\documentclass[a4paper,fleqn,usenatbib]{mnras}

\usepackage{newtxtext,newtxmath}

\usepackage[T1]{fontenc}
\usepackage{ae,aecompl}

\usepackage{graphicx}	
\usepackage{amsmath}	
\usepackage{amssymb}	
\usepackage{gensymb}
\usepackage{physics}
\usepackage{subfigure}
\usepackage{color}
\usepackage{booktabs}
\usepackage[normalem]{ulem}

\definecolor{mygrey}{gray}{0.6}

\title[Radio-Optical Shape and Shear in COSMOS]{Radio-Optical Galaxy Shape and Shear Correlations in the COSMOS Field using 3 GHz VLA Observations}

\author[T. Hillier et al.]{
Tom Hillier$^{1}$\thanks{E-mail: thomas.hillier@manchester.ac.uk},
Michael L. Brown$^{1}$,
Ian Harrison$^{1,2}$
and Lee Whittaker$^{1,3}$
\\
$^{1}$Jodrell Bank Centre for Astrophysics, School of Physics and Astronomy, The University of Manchester, Manchester M13 9PL, UK\\
$^{2}$Department of Physics, University of Oxford, Denys Wilkinson Building, Keble Road, Oxford OX1 3RH, UK\\
$^{3}$Department of Physics and Astronomy, University College London, Gower Street, London WC1E 6BT, UK\\
}

\date{Accepted XXX. Received YYY; in original form ZZZ}

\pubyear{2019}

\begin{document}
\label{firstpage}
\pagerange{\pageref{firstpage}--\pageref{lastpage}}
\maketitle

\begin{abstract}
We present a weak lensing analysis of the 3 GHz VLA radio survey of the COSMOS field, which we correlate with overlapping {\it HST}-ACS optical observations using both intrinsic galaxy shape and cosmic shear correlation statistics. After cross-matching sources between the two catalogues, we measure the correlations of galaxy position angles and find a Pearson correlation coefficient of $0.14 \pm 0.03$. This is a marked improvement from previous studies which found very weak, or non-existent correlations, and gives insight into the emission processes of radio and optical galaxies. We also extract power spectra of averaged galaxy ellipticities (the primary observable for cosmic shear) from the two catalogues, and produce optical-optical, radio-optical and radio-radio spectra. The optical-optical auto-power spectrum was measured to a detection significance of 9.80$\sigma$ and is consistent with previous observations of the same field. For radio spectra (which we do not calibrate, given the unknown nature of their systematics), although we do not detect significant radio-optical (1.50$\sigma$) or radio-radio (1.45$\sigma$) $E$-mode power spectra, we do find the $E$-mode spectra to be more consistent with the shear signal expected from previous studies than with a null signal, and vice versa for $B$-mode and $EB$ cross-correlation spectra. Our results give promise that future radio weak lensing surveys with larger source number densities over larger areas will have the capability to measure significant weak lensing signals.
\end{abstract}




\section{Introduction}
\label{sec:introduction}
Weak gravitational lensing describes the coherent shearing distortion of distant ($z \sim$ 1 and beyond) galaxy shapes due to the curvature of spacetime caused by intervening matter structures. As such, weak lensing is an excellent probe of structure abundance and growth on cosmological scales~\citep[see e.g.][for a review]{Kilbinger2015}.

For the majority of its history, weak lensing has been studied using optical and near infrared wavelengths. Weak lensing due to large scale structure  (or ``cosmic shear") was first detected in the early 2000s~\citep{Bacon2000, vanWaerbeke2000, Wittman2000, Kaiser2000}. The quality and statistical significance of the measurements have steadily increased since then with e.g.~the COMBO-17 survey~\citep{Brown2003a}, the {\it HST}-COSMOS survey~\citep{Massey2007, Schrabback2010}, the  Canada  France  Hawaii  Telescope  Lensing  Survey (CFHTLenS)~\citep[e.g.][]{Heymans2013, Kilbinger2013, Benjamin2013, Kitching2014}, and most recently the Dark Energy Survey (DES)~\citep[e.g.][]{DES2017, Troxel2017}, the Hyper Suprime-Cam (HSC) survey~\citep[e.g.][]{Hikage2019} and the Kilo Degree Survey (KiDS-450)~\citep[e.g.][]{Kohlinger2017}. 

However, future radio surveys also offer interesting prospects for weak lensing studies~\citep{Brown2015, SKAI, SKAII, SKAIII}. This is mainly driven by the advent of the Square Kilometre Array (SKA)~\footnote{\url{ https://www.skatelescope.org/}}, specifically  its mid-frequency dish component (SKA-MID). The SKA telescope will be deployed in two phases, with the first phase (SKA1) providing cosmological constraints comparable to that of Stage III~\footnote{Stages defined by the Dark Energy Task Force (DETF).} optical surveys (e.g. DES and KiDS), while a large cosmic shear survey with the full SKA (SKA2) represents a Stage IV Dark Energy experiment, comparable to the {\it Euclid} and LSST weak lensing surveys.

In addition, radio weak lensing surveys have several unique and powerful features that can potentially help overcome significant outstanding challenges in the field of weak lensing. These features include the use of radio polarisation information to mitigate the effects of intrinsic alignments~\citep{Brown2011}, the precisely known and stable beam response (or Point Spread Function, PSF) of radio interferometers, and the use of radio-optical cross-correlation statistics to eliminate unaccounted-for systematic effects~\citep{Patel2010, Demetroullas2016, SKAIII}. 

The radio telescopes that are most suited to weak lensing studies are interferometers, consisting of arrays of many individual antennae/dishes. Signals between pairs of dishes are correlated to form `visibilities', which essentially sample the Fourier transform of the sky. Pairs of antennae with different projected baselines probe different scales on the sky: larger antennae baselines probe smaller features and smaller baselines probe larger features. To achieve complete Fourier-plane coverage at a single observing frequency, and thus probe all scales on the sky, one would require an infinite number of antennae spread over an infinite distance, tightly packed with no gaps or shadowing. However, broad frequency bandwidth coverage probes a range of different spatial scales for sources with well-determined spectra and may be used (along with large numbers of telescopes) to fill out the Fourier-plane coverage to a high fraction. This is the driving force behind the SKA. In particular, the smallest spatial scales measurable with the SKA will be $\sim0.5$ arcsec (for SKA Band 2, centred at $\nu = 1.355$ GHz). This will facilitate shape measurements for large numbers of $z\sim1$ galaxies which will be detected in deep and wide area surveys, crucial for weak lensing studies. This capability is primarily due to the increased number and range of baselines in the SKA: 194 dishes and a longest baseline of $\sim150$ km for SKA1, and $\sim2000$ dishes with baselines extending over $1000$ km for SKA2.

The SKA will be a major advance compared to existing radio facilities. In the context of weak lensing, the most relevant existing radio interferometers are the National Science Foundation's Karl G. Jansky Very Large Array (VLA) in the USA and the e-MERLIN telescope in the UK. The VLA, data from which we use in this work, is well suited to weak lensing due to its combination of excellent point source sensitivity and smallest measurable angular scale. The VLA has a total of 27 dishes, and a longest baseline of 36 km. The e-MERLIN telescope is a 7-element interferometer, with a longest baseline of 217 km. While not as sensitive as the VLA, e-MERLIN does provide additional information on finer angular scales, which can also be important for weak lensing. Indeed, the combination of data from the VLA and e-MERLIN has been used in the past to perform detailed morphological studies of high-redshift galaxies~\citep{Muxlow2005}. The combination of VLA and e-MERLIN observations is also a driving feature of the ongoing SuperCLASS survey~\footnote{\url{www.e-merlin.ac.uk/legacy/projects/superclass.html}}, which aims to detect the radio weak lensing signal from a supercluster system.

The current major advantage of optical surveys over radio surveys for weak lensing studies is the much larger source number densities typically achieved in the former -- weak lensing requires a large source number density over a large area. For instance, leading optical surveys have source number densities of $\sim10$ galaxies arcmin$^{-2}$ over several hundred square degrees~\citep[e.g.][]{Hildebrandt2017}, whereas even the deepest radio surveys (including the survey we analyse here) typically have number densities of $\sim1$ galaxy arcmin$^{-2}$ over only a few square degrees~\citep[e.g.][]{Schinnerer2010, Smolcic2017}.

In a previous related work,~\cite{Tunbridge2016} used archival L-band VLA observations~\citep[centred at $1.4$ GHz,][]{Schinnerer2010} of the Cosmic Evolution Survey (COSMOS, \citealt{Scoville2007}) field to assess the suitability of existing radio data for measuring the shapes of distant star-forming galaxies (SFGs), and thus their use for weak lensing studies. Since these observations were conducted at L-band (and with the VLA primarily in its most extended ``A-array" configuration), the effective angular resolution of the radio image used in that study was $\sim1.5$ arcsec, which limited the ability to extract reliable galaxy shapes for large numbers of star-forming galaxies.

In this work, we make use of more recent S-band observations of the COSMOS field \citep[centred at 3 GHz,][]{Smolcic2017} to perform an exploratory radio weak lensing analysis. These observations were also obtained mainly using the VLA's most extended A-array configuration. However, because the angular scale probed by any given baseline scales inversely with observing frequency, the effective angular resolution of the 3 GHz data is $\sim0.75$ arcsec. This is much better suited to extracting accurate galaxy shape measurements (with $z\sim1$ galaxies typically having sizes $\sim1$ arcsec) and is comparable to the angular resolution achieved in leading ground-based optical lensing surveys (c.f. median seeing conditions of 0.96 arcsec for DES, \citealt{Zuntz2017}, and 0.66 arcsec for KiDS, \citealt{Hildebrandt2017}).

In addition to demonstrating precise galaxy shape measurements in the radio, these data offer an opportunity to explore further the degree to which the shapes of galaxies, as observed in the radio and optical bands, are correlated. Such measurements can provide insight into the physical processes associated with each waveband. The emission processes of SFGs for radio and optical wavelengths are different. For optical wavelengths, the emission is dominated by the star-forming regions of the galaxy. In contrast, for radio wavelengths, the emission is dominated by synchrotron radiation. It is expected that these two processes should trace each other, and therefore should give similar estimates of the galaxy shape.

In addition to SFGs, radio surveys are also sensitive to Active Galactic Nuclei (AGN), which typically have their strongest radio emission anti-aligned with the main disk structure of the galaxy. The relative abundances of AGN and SFGs detected in a radio survey are heavily dependent on the survey flux limit, with shallow surveys being dominated by AGN and deep surveys being dominated by SFGs. 

Radio-optical cross-correlation galaxy shape analyses have been attempted before with varying results. \cite{Patel2010} cross-correlated observations from the VLA and MERLIN radio surveys with optical observations from the {\it HST} Great Observatories Origins Deep Survey (GOODS). In this analysis, no correlation was found between the intrinsic ellipticities of matched objects found in both the optical and radio surveys. The previous analysis~\citep[][including some of the authors]{Tunbridge2016} also failed to detect a significant radio-optical shape correlation, although it did place an upper limit on the allowed strength of such a correlation. \cite{Battye2009} did detect a correlation between the shapes of galaxies as measured in the radio and optical bands using the Sloan Digital Sky Survey (SDSS) and Faint Images of the Radio Sky at Twenty centimetres (FIRST) survey data. However, their sample is composed of nearby, well-resolved galaxies only and -- unlike the sample analysed here -- is not representative of the high-redshift SFG population expected to dominate future deep radio surveys.

The 3 GHz COSMOS data is also the premier existing dataset for demonstrating the power of radio-optical cross-correlation weak lensing statistics. In particular, the existence of overlapping deep {\it Hubble Space Telescope} (\textit{HST}) observations make the COSMOS field an ideal testbed for extracting the radio-optical cross-correlation power spectrum. \cite{Demetroullas2016} presented a cross-correlation cosmic shear power spectrum analysis of the SDSS and VLA-FIRST surveys. This analysis measured a cosmic shear signal at a significance of $\sim2.7\sigma$, while also demonstrating the removal of wavelength-dependent systematics through cross-correlating data. However, once again, the sample analysed is not representative of future radio surveys, the majority of sources in the shallow VLA-FIRST survey being AGN.  

This paper is organised as follows. In Section~\ref{sec:theory} we present the relevant weak lensing theory and the background material for our power spectrum analysis. In Section~\ref{sec:data}, we summarize the key aspects of the datasets we have used and we discuss the potential impact of the differing projections used in the creation of the {\it HST} and VLA images. Section~\ref{sec:shapes} describes our shape measurements and Section~\ref{sec:sourceselection} discusses how we have selected sources for the subsequent analyses. In Section~\ref{sec:shapecomparison}, we present a comparison of the intrinsic radio and optical shapes of those galaxies that are detected in both wavebands. The main results of our analysis -- the cosmic shear power spectra -- are presented in Section~\ref{sec:powerspectra}. We conclude in Section~\ref{sec:conclusions}.

\section{Theory}
\label{sec:theory}

\subsection{Weak Lensing Theory}
For a thorough introduction to weak lensing, we refer the reader to \cite{Bartelmann2001} and \cite{Kilbinger2015}. Here, we simply summarize the relevant quantities for the analysis that follows. 

The shapes of SFGs can be modelled as ellipses, which can be parameterised using their two-component ellipticity, $\epsilon_i$, for $i = \{1,2\}$. Here, $\epsilon_1$ parametrises elongations parallel and perpendicular to an arbitrary Cartesian reference axis, and $\epsilon_2$ describes elongations at $\pm$45$\degree$ relative to the same reference axis. One can also completely specify the shape of an ellipse using a position angle, $\alpha_{\rm{p}}$, a semi-major axis, $a$ and a semi-minor axis, $b$. The ellipticity components are related to these latter quantities by
\begin{align}
&\epsilon_1=|\epsilon| \cos(2\alpha_{\rm{p}}), \nonumber \\
&\epsilon_2=|\epsilon| \sin(2\alpha_{\rm{p}}),
\end{align}
where the ellipticity modulus, $|\epsilon|$ is
\begin{equation}
	|\epsilon| = \frac{a^2 - b^2}{a^2 + b^2} = \sqrt{\epsilon_1^2 + \epsilon_2^2}
	\label{eq:ellipticity}.
\end{equation}

Intervening large-scale structure has the effect of distorting the shapes of background galaxies. Such a distortion can be described by the two-component shear, $\gamma_i$, defined in a similar way to the ellipticity above. The shear quantifies the extent to which the ellipticity of a source galaxy, $\epsilon^{\rm{s}}_i$ is altered to give the observed image ellipticity, $\epsilon_i$. To first order, this alteration is described by 
\begin{equation}
	\epsilon_i = \epsilon_i^{\rm{s}} + \gamma_i
	\label{eq:ellipticityshear},
\end{equation}
where $\gamma_i$ is specific to each galaxy image. The shear (and ellipticity) can also be written as a complex number,
\begin{equation}
	\gamma \equiv \gamma_1 + i \gamma_2  = \abs{\gamma} e^{2 i \phi}
	\label{eq:shearbasic},
\end{equation}
where $\abs{\gamma} = \sqrt{\gamma_1^2 + \gamma_2^2}$. $\abs{\gamma}$ and $\phi$ describe the magnitude and orientation of the distortion respectively, where $\abs{\gamma} \ll 1$ in the weak lensing regime. The factor of 2 in the phase is a result of the rotational symmetry of ellipses: an ellipse maps onto itself after a rotation of 180$\degree$. Thus the shear (or ellipticity) is known as a ``spin-2" field. 

If one assumes that the source galaxy shapes are randomly distributed, i.e.
\begin{equation}
	\left < \epsilon^{\rm{s}} \right > = 0
	\label{eq:randomdistassumption},
\end{equation}
where the angled brackets denote an expectation value, then equation~(\ref{eq:ellipticityshear}) suggests that the shear in a sufficiently small region of sky can be estimated using
\begin{equation}
	\widehat{\gamma} = \frac{1}{\sum_i w_i} \sum_i w_i \epsilon_i
	\label{eq:avg_shear},
\end{equation}
where $w_i$ is an arbitrary weighting and the sums extend over all galaxies in the region of interest. Though we do not do so here, one can also estimate the projected surface mass density, or lensing convergence field, $\kappa$ (a scalar, or ``spin-0" field), from the observed magnitudes and/or sizes of background galaxies~\citep[e.g.][]{Alsing2015}. 

The shear and convergence can be written in terms of a 2D lensing potential, $\psi$ as
\begin{equation}
	\gamma_1 = \frac{1}{2} (\partial_1\partial_1 - \partial_2\partial_2) \,\psi, \,\,\,\, \gamma_2 = \partial_1\partial_2 \,\psi, \,\,\,\, \kappa = \frac{1}{2} (\partial_1\partial_1 + \partial_2\partial_2) \,\psi
	\label{eq:shearrelations},
\end{equation}
where $\partial_i$ denote partial derivatives with respect to Cartesian coordinates, $\theta_i$. In turn the lensing potential is related to the 3D gravitational potential $\Phi(\mathbf{r})$,  by~\citep{Kaiser1998}
\begin{equation}
	\psi(\boldsymbol{\theta}) = \frac{2}{c^2} \int_0^r dr' \left(\frac{r-r'}{rr'}\right) \Phi(\mathbf{r'}),
\end{equation}
where $\boldsymbol{\theta}$ is the angular position on the sky, $c$ is the speed of light in a vacuum, $\mathbf{r}$ is the 3D comoving position, and $r$ is the comoving distance to the source galaxies.

It is often useful to decompose the shear field into a gradient and a curl component, referred to as $E$- and $B$-modes respectively. Weak lensing is produced by scalar perturbations in spacetime. Therefore, it is expected that the shear field contains no handedness. Hence, the expected weak lensing signal is a pure gradient ($E$-mode) component, while the curl ($B$-mode) component is expected to be zero. $B$-mode signals can be produced by gravitational waves \cite[e.g.][]{Dodelson2003} or by clustering of source galaxies~\cite[][]{Schneider2002} but both of these effects are expected to be negligible. More likely causes for the generation of $B$-modes at detectable levels are residual instrumental/observational systematic effects or intrinsic alignments of galaxy shapes~\citep[e.g.][]{Heavens2000, Catelan2001, Crittenden2001, Brown2002}. 

\subsection{Shear Power Spectra}
Statistical measurements of weak lensing typically employ two-point estimators such as the real-space correlation function or its Fourier-space analog, the power spectrum, which we use here. In this work, in addition to estimating the (auto-) shear power spectra in the optical and radio wavebands separately, we also estimate the cross-correlation shear power between the two wavebands. 

We consider the lensing (convergence) power spectra, $C_\ell^{\kappa(i)\,\kappa(j)}$, where the indices $i$ and $j$ represent different galaxy samples. In this analysis, $i$ and $j$ can each take values R (for radio) or O (for optical), giving three possible spectra. The power spectra are directly related to the 3D matter power spectrum, $P_\delta(k)$ through~\cite[e.g.][]{Bartelmann2001}
\begin{equation}
	C_\ell^{\kappa(i)\,\kappa(j)} = \frac{9 H_0^4 \Omega_{\mathrm{m}}^2}{4 c^4} \int_{0}^{\chi_{\mathrm{h}}} d\chi \frac{g^i(\chi)\,g^j(\chi)}{a^2(\chi)} P_\delta \left(\frac{\ell}{f_K (\chi)}, \chi \right),
    \label{eq:powspecfull}
\end{equation}
where $H_0$ is the Hubble constant, $\Omega_\mathrm{m}$ is the total matter density, $c$ is the speed of light in a vacuum, $\chi$ is the co-moving distance, $a (\chi)$ is the scale factor of the Universe and $f_K (\chi)$ is the angular diameter distance ($f_K (\chi) = \chi$ for a flat Universe). The kernels, $g^{i, j} (\chi)$ describe the relative contributions of the two galaxy samples to the lensing signal, and are given by
\begin{equation}
	g^{i} (\chi) = \int_{\chi}^{\chi_{\mathrm{h}}} d\chi^\prime n_{i} (\chi^\prime) \frac{f_K (\chi^\prime - \chi)}{f_K (\chi^\prime)},
    \label{eq:lensing kernels}
\end{equation}
where $n_i (\chi)$ is the distribution of galaxies, as a function of comoving distance, for galaxy sample $i$, and the integral extends to the horizon, $\chi_{\mathrm{h}}$. In the limit of weak lensing, the two-point statistical properties of the shear and convergence fields are the same~\citep{Blandford1991}, so that 
\begin{equation}
C_\ell^{\gamma(i)\,\gamma(j)} = C_\ell^{\kappa(i)\,\kappa(j)}.
\label{eq:shear_conv_equiv}
\end{equation}		
For the remainder of this paper, we will refer only to shear power spectra and so we will use a more concise notation for the three possible spectra: $C_\ell^{\rm RR}, C_\ell^{\rm OO}$ and $C_\ell^{\rm RO}$, e.g. $C_\ell^{\rm RO} \equiv C_\ell^{\gamma({\rm R})\,\gamma({\rm O})}$.

\subsection{Cross-correlation noise terms}
\label{sec:crosscorrelationnoise}
When estimating the shear power spectra from real data, one needs to account for the intrinsic galaxy shape noise contribution to the measured ellipticity correlations. This can be done by specifying the power spectrum of the noise, $\mathcal{N}^{ij}(\ell)$ for each galaxy sample combination $\{i, j\}$. In the usual case, where the galaxy samples are from two tomographic bins from the same experiment, the noise power spectrum is simply
\begin{equation}
	\mathcal{N}^{i j} (\ell) = \delta_{i j} \frac{\sigma_{\epsilon, i}^2}{n_{\mathrm{gal}}^i},
    \label{eq:noiseauto}
\end{equation}
where $\delta_{ij}$ is the Kronecker delta function, $n^i_{\rm gal}$ is the number density of galaxies in the $i^{\rm th}$ galaxy sample, and $\sigma_{\epsilon, i}$ is the dispersion of the galaxy intrinsic ellipticities in the $i^{\rm th}$ sample. 

However, this term can be more complicated if the two galaxy samples contain sources common to both, as could be the case in a radio-optical cross-correlation analysis. In this case (setting $i = {\rm R}$ and $j = {\rm O}$), the noise power spectrum is given by~\citep{SKAI}:
\begin{equation}
	\begin{split}
	\mathcal{N}^{\rm RO}(\ell) =& \frac{1}{n_{\mathrm{gal}}^{\rm R} n_{\mathrm{gal}}^{\rm O}} \big \langle \sum_{\alpha \in {\rm R}} \epsilon_\alpha \sum_{\beta \in {\rm O}} \epsilon_\beta \big \rangle \\
	=& \frac{n_{\rm gal}^{\rm RO}}{n^{\rm R}_{\rm gal}n^{\rm O}_{\rm gal}} {\rm cov}(\epsilon_{\rm R}, \epsilon_{\rm O}).
    \end{split}
    \label{eq:noisecrossfull}
\end{equation}
Here, $n_{\rm gal}^{\rm RO}$ is the number density of galaxies that are common to both samples and ${\rm cov}(\epsilon_{\rm R}, \epsilon_{\rm O})$ is the covariance of the intrinsic shapes of those common galaxies in the radio and optical wavebands. If the shapes of galaxies in the optical and radio wavebands are identical, then ${\rm cov}(\epsilon_{\rm R}, \epsilon_{\rm O}) = \sigma^2_{\epsilon, {\rm R}} = \sigma^2_{\epsilon, {\rm O}}$, while if they are completely uncorrelated, this term is zero. The strength of such radio-optical intrinsic shape correlations in the real Universe is very uncertain at present (see the discussion in Section~\ref{sec:introduction}) and is something that we investigate with the COSMOS data in Section~\ref{sec:shapecomparison}. 

\section{Data}
\label{sec:data}

\subsection{The COSMOS Field}
The surveys analysed in this work cover the COSMOS field~\footnote{\url{ http://cosmos.astro.caltech.edu/}}, a $\sim2 \deg^2$ patch of sky centred at (RA, Dec) = ($150 \deg, +2 \deg$). This region has been surveyed with many telescopes across a wide range of wavelengths including {\it Chandra} (X-ray), {\it Spitzer} (infrared), {\it HST} (optical), {\it GALEX} (UV) and the Very Large Array (VLA, radio). Hence there is a plethora of datasets that can be cross-matched. In this work, we cross-match the 3 GHz VLA data with \textit{HST}-ACS optical data. More details on these observations are given in Sections~\ref{sec:radio_observations} and~\ref{sec:optical_observations}.

\subsection{VLA Radio Observations}
\label{sec:radio_observations}
The VLA 3 GHz observations of the COSMOS field were collected as described in~\cite{Smolcic2017}. The data were collected over a total of 384 hours in the VLA's A-array (324 hours) and C-array (60 hours) configurations, and covered a bandwidth of 2048 MHz, centred on 3 GHz. 

A total of 192 pointings were observed over an area of $2.6 \deg^2$, covering the full COSMOS field. The layout of these pointings followed a square grid containing 64 pointings evenly spaced by 10$^{\prime}$ in Right Ascension (RA) and Declination (Dec), which corresponded to two-thirds of the half-power beam width (HPBW) at the central frequency of 3~GHz. Two further sets of 64 pointings were added to this: one shifted by +5$^{\prime}$ in RA and Dec and the other shifted by -5$^{\prime}$ in RA only. This was done to achieve a uniform root-mean-square (rms) noise level across the whole field.

\subsubsection{Image Creation}
\label{sec:imagecreation}
The real-space image used for this work was provided by V. Smol\v{c}i\'{c} and M. Novak via private communication. A full description of the imaging procedure is presented in~\citet{Smolcic2017}, and the image is now publicly available from \url{http://jvla-cosmos.phy.hr/Home.html}.

The image was created using the multiscale multifrequency synthesis (MSMF) algorithm~\citep{Rau2011} implemented in \textsc{casa}. MSMF simultaneously uses the whole 2 GHz bandwidth to calculate the monochromatic flux density at 3 GHz and a spectral index for frequencies between 2 and 4 GHz. It was found that the MSMF produced the best resolution, rms and image quality in comparison to other imaging methods~\citep{Novak2015}.

Each pointing was imaged individually and combined in the image plane to produce the full mosaic. This was done in the image plane primarily because of the large data volume, making joint deconvolution impractical. The mosaic was \textsc{clean}ed down to 5$\sigma$ and then visually inspected to manually define masks (to block out bright sources) for further \textsc{clean}ing down to 1.5$\sigma$. The final image mosaic has a median rms noise level of 2.3 $\mu$Jy beam$^{-1}$ over the full $2 \deg^2$ COSMOS field.

Synthesised beam sizes between pointings were found to vary by $\approx$ 0.03$^{\prime \prime}$. These variations were deemed small enough to allow the PSF of the image to be modelled using an average circular Gaussian beam with a Full Width at Half Maximum (FWHM) of $0.75^{\prime \prime}$.

\subsection{\textit{HST}-ACS Optical Survey Data}
\label{sec:optical_observations}
The optical observations used for cross-matching were collected using the \textit{HST}-Advanced Camera for Surveys (\textit{HST}-ACS) over a total of 583 \textit{HST} orbits, covering a field area of $1.64 \deg^2$, as described in \cite{Koekemoer2007}.

The observations used the F814W filter and reach a limiting magnitude of $I_{\mathrm{AB}} = 27.2$ mag at $10\sigma$, with an angular resolution of $0.05^{\prime \prime}$~\citep{McCracken2010}. The science data used in this study was produced by~\cite{Koekemoer2007} and consists of 575 pointings across the entire mosaic. Additionally, for each pointing the raw exposures were combined using the MultiDrizzle pipeline~\citep{Koekemoer2003} to improve the pixel scale to 0.03$^{\prime \prime}$. The {\it HST}-ACS image that we use here is the same as that used for the analysis presented in~\cite{Tunbridge2016}. The output MultiDrizzled image cut-outs are available at \url{http://irsa.ipac.caltech.edu/Missions/cosmos.html}.

\subsection{Coordinate Projections}
\label{sec:coord_projections}
Creating a 2D image from a set of observations requires projecting the right-ascension, $\alpha$ and declination, $\delta$ sky coordinates onto flat-plane, Cartesian coordinates, $x$ and $y$. There are several different projections that can be used to do this. Two of the most common are the gnomonic/tangential and orthographic projections. The 3 GHz VLA radio image was created using an orthographic projection and the \textit{HST}-ACS optical image was created using a gnomonic projection. The gnomonic projection has its projection point at the centre of a great circle (the position of the telescope) and the orthographic projection has its projection point at infinity.

Here we demonstrate that, for the small survey area of COSMOS, these two projections are approximately the same, thus facilitating a direct comparison of the shapes of the sources as measured from the two images. The argument that follows is almost identical to the justification for the flat-sky approximation, whereby a curved sky can be approximated to be flat for surveys spanning $\lesssim 10 \degree$. Clearly, this approximation holds for our two datasets, but we also directly compare the two different projections here to justify using the two images in comparison to one another.

The two mappings from sky to Cartesian coordinates are given by
\begin{equation}
	\begin{split}
	x_{\mathrm{gnomonic}} &= \frac{\cos\delta \sin(\alpha - \alpha_0)}{\cos c} \\
    y_{\mathrm{gnomonic}} &= \frac{\cos\delta_0 \sin\delta - \sin\delta_0 \cos\delta \cos(\alpha - \alpha_0)}{\cos c}
	\end{split}
    \label{eq:projections_gnomonic}
\end{equation}
and
\begin{equation}
	\begin{split}
	x_{\mathrm{orthographic}} &= \cos\delta \sin(\alpha - \alpha_0) \\
    &\equiv x_{\mathrm{gnomonic}} \cos c \\
    y_{\mathrm{orthographic}} &= \cos\delta_0 \sin\delta - \sin\delta_0 \cos\delta \cos(\alpha - \alpha_0) \\
    &\equiv y_{\mathrm{gnomonic}} \cos c
    \end{split}
	\label{eq:projections_orthographic}
\end{equation}
where ($\alpha_0$, $\delta_0$) is the centre of projection on the sky. The translation term is given by
\begin{equation}
    \cos c = \sin\delta_0 \sin\delta + \cos\delta_0 \cos\delta \cos(\alpha - \alpha_0) \, \mathrm{.}
	\label{eq:projections_c}
\end{equation}
As can be seen from this equation, the translation term between an orthographic and gnomonic projection is a function of the sky coordinates ($\alpha, \delta$) and the projection centre ($\alpha_0, \delta_0$). In the case of small survey angles ($\alpha \approx \alpha_0$ and $\delta \approx \delta_0$) and small deviations between the projection centres for each map ($\alpha_0^{\mathrm{optical}} \approx \alpha_0^{\mathrm{radio}}$ and $\delta_0^{\mathrm{optical}} \approx \delta_0^{\mathrm{radio}}$) equation~(\ref{eq:projections_c}) approximates to $\cos c \approx 1$ and the two mappings are the same.

These approximations hold for the surveys in this work since the COSMOS field only spans an angle of $\sim \sqrt[]{2} \degree$, and both the 3 GHz VLA radio and \textit{HST}-ACS optical surveys are centred on the centre of the field.

\section{Shape Measurement}
\label{sec:shapes}
To measure the shapes of galaxies, we use~\textsc{im3shape}~\citep{Zuntz2013}, which is a maximum likelihood model fitting algorithm, and is one of two shape measurement algorithms used in the DES weak lensing analyses~\citep{Zuntz2017}. \textsc{im3shape} takes an input image, source positions, and a model of the PSF in the image, and fits a brightness distribution to the galaxy shapes. We choose the brightness distribution of galaxies to be described by S\'{e}rsic profiles:
\begin{equation}
	I(r) = I(0) \exp \left[ - \left( \frac{r}{r_0} \right)^{\frac{1}{n}} \right ] ,
	\label{eq:sersic_profile}
\end{equation}
where $I(0)$ is the central intensity, $r$ is the radius perpendicular to the line of sight, $r_0$ is the galaxy scale length, and $n$ is the S\'{e}rsic index.

\subsection{Radio Shapes}
\label{sec:radioshapes}
To measure the galaxy shapes, \textsc{im3shape} was applied to the real-space image introduced in Section~\ref{sec:imagecreation}. As mentioned previously the PSF of the image can be modelled using a circular Gaussian beam of $0.75^{\prime \prime}$. Hence, we provide \textsc{im3shape} with a circular Gaussian PSF, with a FWHM of $0.75^{\prime \prime}$. We use a two-component S\'{e}rsic bulge$+$disc galaxy model, which is the sum of two S\'{e}rsic profiles, equation~(\ref{eq:sersic_profile}). More specifically, we use a de Vaucouleurs bulge ($n=4$) and an exponential disc ($n=1$) model~\citep{Zuntz2013}.

Ideally, galaxy shapes from radio observations would be measured in Fourier-space (the space in which the data are taken). The method we use here -- measuring radio shapes from a real-space image -- is not optimal and does not fully utilize one of the main advantages of using radio observations, which is that the PSF should be exactly known from the positions of the antennae, as mentioned in Section~\ref{sec:introduction}. It also introduces sources of systematic error due to (i) the nonlinear deconvolution step that occurs during imaging and (ii) the correlated noise introduced into the image by the side-lobes produced by every source in the observation.

Fourier-plane shape extraction methods can, in principle, circumvent these issues but they are currently less advanced than the process of radio imaging followed by real-space shape fitting. More sophisticated methods for Fourier-plane shape extraction are being developed for the SKA~\citep[e.g.][]{Rivi2016, Rivi2018, Rivi2018a}, and as part of the SuperCLASS project. 

However, we stress that shape measurements derived from the image plane are sufficient for the analysis presented here, given the low number density of radio galaxies that are retained in the final weak lensing analysis. As detailed in Section~\ref{sec:powerspectra}, the errors in our measured radio-radio and radio-optical shear power spectra are dominated by noise due to the low source number density of the radio weak lensing catalogue, and are much larger than any systematic errors introduced by the shape measurement procedure. 

\subsection{Optical Shapes}
The optical galaxy shape catalogue was created as described in~\cite{Tunbridge2016} using the \textit{HST}-ACS data described in Section~\ref{sec:optical_observations}. This made use of several software packages including the source extraction software, SE\textsc{xtractor}~\citep{Bertin1996} as used in previous weak lensing analyses~\citep[e.g.][]{Jarvis2016}, and the PSF extractor software, \textsc{psfex}~\citep{Bertin2011}.

The shapes for this catalogue were also measured using \textsc{im3shape} with a two-component S\'{e}rsic bulge$+$disc galaxy model. We refer the reader to \cite{Tunbridge2016} for further details on the construction of the optical shape catalogue. 

\section{Source Selection}
\label{sec:sourceselection}
To carry out a weak lensing analysis, the sources in the catalogues should ideally have similar morphologies, such that a consistent model can be fitted to each of them. Any regular source morphology would satisfy these criteria. However, it is expected that the galaxy populations probed by deep optical and radio observations will be dominated by SFGs with elliptical morphologies. In addition, since most of the current analysis infrastructure is based around this assumption, these SFGs were selected for the lensing shear analysis.

The initial 3 GHz VLA radio and \textit{HST}-ACS optical source catalogues contained 10,830 and 438,226 sources respectively. To remove unwanted sources from these catalogues including stars (optical) and unresolved sources, several selection criteria were applied. Firstly, in both the 3~GHz radio and \textit{HST}-ACS optical catalogues, sources measured to have radii smaller than half the PSF FWHM were classified as unresolved and removed from the catalogues. This left 3,759 and 299,806 sources in the radio and optical catalogues. Further selection criteria, specific to each catalogue are discussed in the next two sections and a summary of all the cuts made is given in Table~\ref{tab:WLSourceSelection}.

\begin{table}
	\centering
	\caption{Summary of the cuts made to each catalogue to form the weak lensing catalogues. The cuts are listed in the order they were applied and the values and percentages quantify the number of sources remaining after each cut.}
	\label{tab:WLSourceSelection}
	\begin{tabular}{lcc} 
		\hline
		Cut Description & 3 GHz VLA & \textit{HST}-ACS \\
		\hline
		Full Catalogue & 10,830 (100\%) & 438,226 (100\%) \\
		Radius $>$ 0.5 $\times$ PSF FWHM & 3,759 (35\%) & 299,806 (68\%) \\
		Remove Stars & - & 297,432 (68\%) \\
		i-band Mask & - & 245,865 (56\%) \\
		Ellipticity Cuts (see text) & 2,038 (19\%) & 243,852 (56\%) \\
		\hline
	\end{tabular}
\end{table}

\subsection{3 GHz VLA Radio Weak Lensing Catalogue}
For the 3 GHz radio catalogue, a very large excess of very low ellipticity (near-circular) sources were found with $| \epsilon | \le 0.05$. Further investigation revealed no obvious correlation of these low-ellipticity objects with any other source properties including galaxy size, location, SNR, flux density and/or the local rms noise level. We therefore cut our catalogue based on this ellipticity, removing sources with $| \epsilon | \le 0.05$, on the understanding that this minimum retained ellipticity is small enough that our shear power spectrum constraints will not be significantly biased (though the level of noise will be changed due to the decrease in the number of sources). The final weak lensing catalogue for the radio data contained 2,038 sources.

\subsection{\textit{HST}-ACS Optical Weak Lensing Catalogue}
\begin{figure}
	\includegraphics[width=\columnwidth]{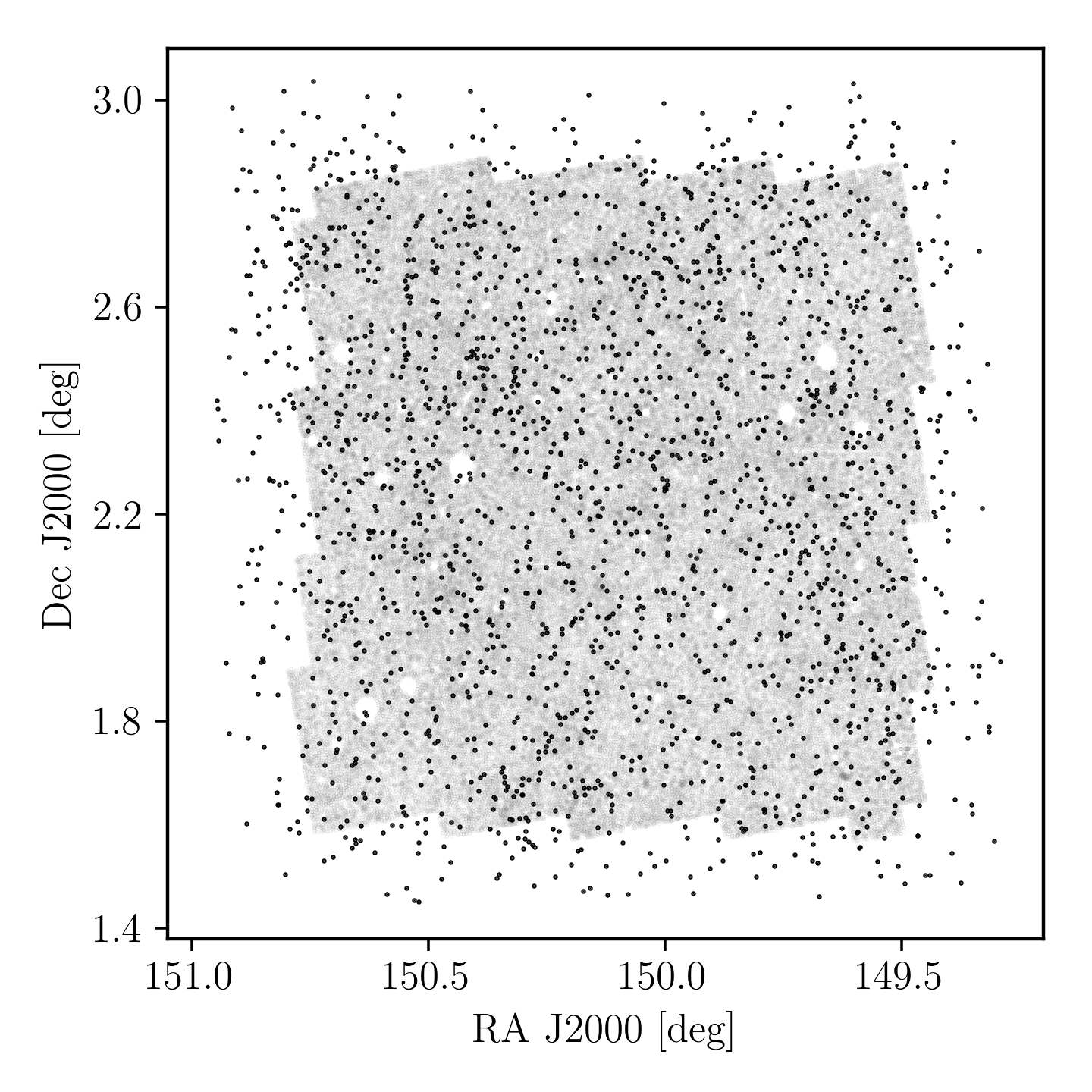}
    \caption{The distribution of sources selected for the weak lensing analysis: the {\it HST}-ACS optical source positions (grey) are overlaid with the 3 GHz VLA radio source positions (black). The radio points are plotted with markers 200 times the size of the markers used for the optical points.}
    \label{fig:ScatterMap_afterCuts}
\end{figure}

For the {\it HST}-ACS optical catalogue, we use the publicly available COSMOS source catalogue~\footnote{\url{http://irsa.ipac.caltech.edu/data/COSMOS/}} created as described in~\cite{Capak2007} to remove all objects classified as stars, leaving 297,432 sources, and to apply an i-band mask to remove areas where the camera CCDs were being saturated, leaving 245,865 sources. The effect of this mask can be seen in Fig.~\ref{fig:ScatterMap_afterCuts}, where the empty circles in the {\it HST} coverage are the locations of charge bleeds from saturated stars, and the small streaks are bleeds along CCD boundaries.

\textsc{im3shape} returned a slight excess of sources with high ellipticies, $|\epsilon| \ge 0.9$. On further inspection, these sources were found to be associated with a poor goodness of fit, and so were removed from the shape catalogue. After these cuts, the final optical shape catalogue contained 243,852 sources.

\subsection{Weak Lensing Diagnostics}
One diagnostic we have used to assess the quality of the weak lensing catalogues is the distributions of ellipticity moduli. Following equation~(\ref{eq:ellipticity}), and assuming $\epsilon_1$ and $\epsilon_2$ to be normally distributed, the quadrature sum, $| \epsilon | = \sqrt[]{\epsilon_1^2 + \epsilon_2^2}$ should follow a Rayleigh distribution. More generally, the distribution of measured ellipticities is often modelled using parameterised fitting formuale. Here, we chose to use the empirical formula \citep{Bridle2010},
		\begin{equation}
			P(| \epsilon |) = \epsilon \left [ \cos \left ( \frac{\pi \epsilon}{2} \right ) \right ] ^2 \exp \left [ - \left ( \frac{2 \epsilon}{B} \right ) ^C \right ]
		\label{eq:ellipticity_modulus_fit},
		\end{equation}
	where $B$ and $C$ are constants. This equation was loosely motivated by results from the APM survey \citep{Crittenden2001} and gives a good indication of the general expected shape of the distribution. 
    
\begin{figure}
	\includegraphics[width=\columnwidth]{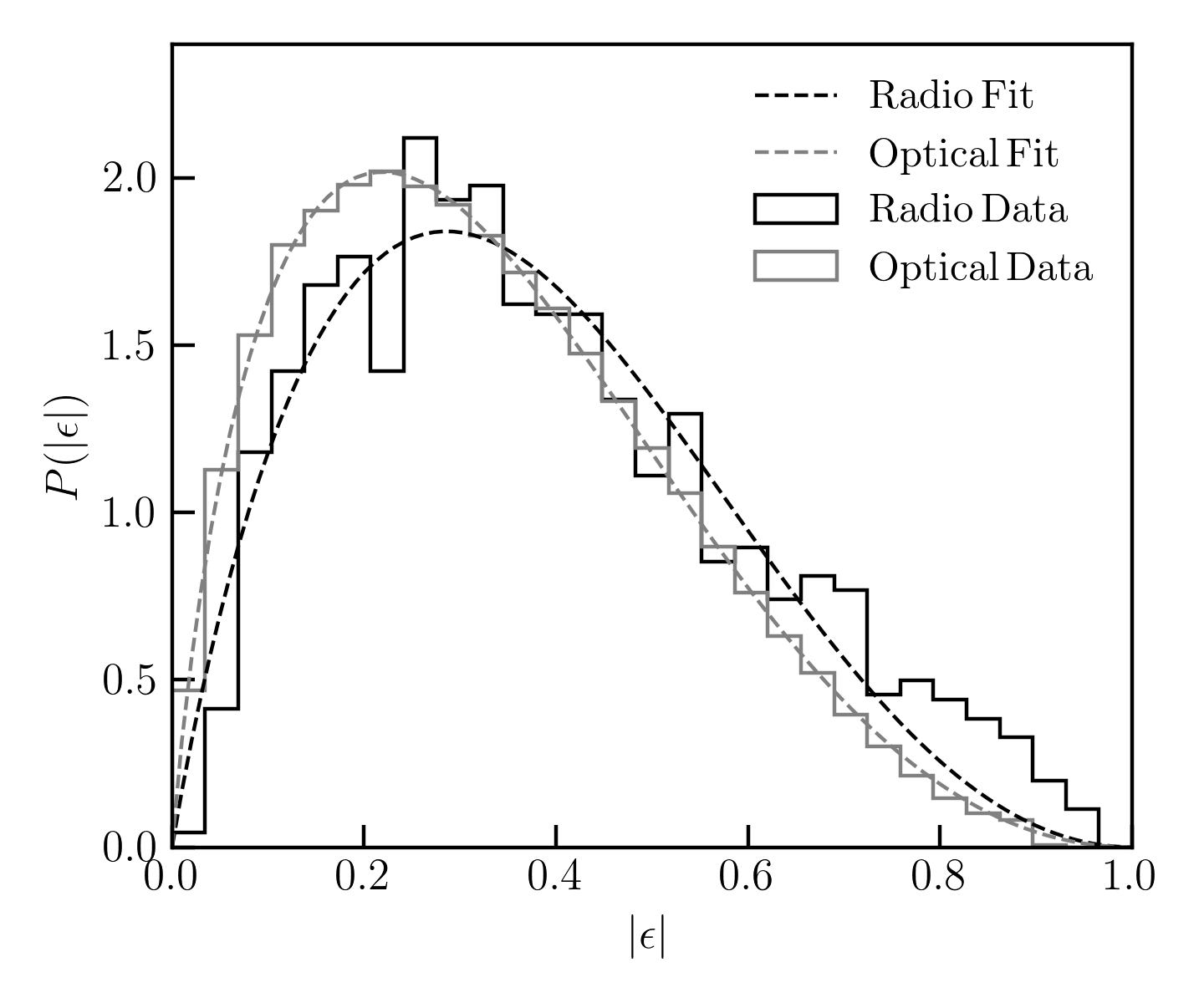}
    \caption{Normalised ellipticity moduli distributions for the 3 GHz VLA radio and {\it HST}-ACS optical weak lensing catalogues. The dashed curves show fits to the data using the empirical formula given in equation~(\ref{eq:ellipticity_modulus_fit}).}
    \label{fig:hist_ellipticity_both}
\end{figure}

Fig.~\ref{fig:hist_ellipticity_both} shows the ellipticity moduli distribution of the 3 GHz VLA radio and of the {\it HST}-ACS sources selected for weak lensing. The data distributions are reasonably well fit using the model distribution given in equation~(\ref{eq:ellipticity_modulus_fit}), as indicated by the dashed curves. For the radio dataset, we measure best-fitting model parameter values of $B = 0.78 \pm 0.42$ and $C = 0.71 \pm 0.31$ and for the optical dataset, we measure $B = 0.27 \pm 0.02$ and $C = 0.57 \pm 0.01$. For reference, the representative model used in the GREAT08 analysis of~\cite{Bridle2010} had parameter values of $B = 0.19$ and $C = 0.58$.

Another useful diagnostic is the distribution of the galaxy position angles, 
\begin{equation}
	\alpha_{\rm{p}} = \frac{1}{2} \tan^{-1} \left ( \frac{\epsilon_2}{\epsilon_1} \right )
	\label{eq:posangle},
\end{equation}
where the possible range of $\alpha_{\rm{p}}$ values is $0\degree \le \alpha_{\rm{p}} \le 180\degree$, covering all possible orientations of an ellipse. For a large enough sample, the distribution of position angles should be consistent with a uniform distribution. As such, it is a good test of homogeneity, and any deviations from uniform could indicate an issue with the analysis methods, or the presence of intrinsic alignments.

\begin{figure}
	\includegraphics[width=\columnwidth]{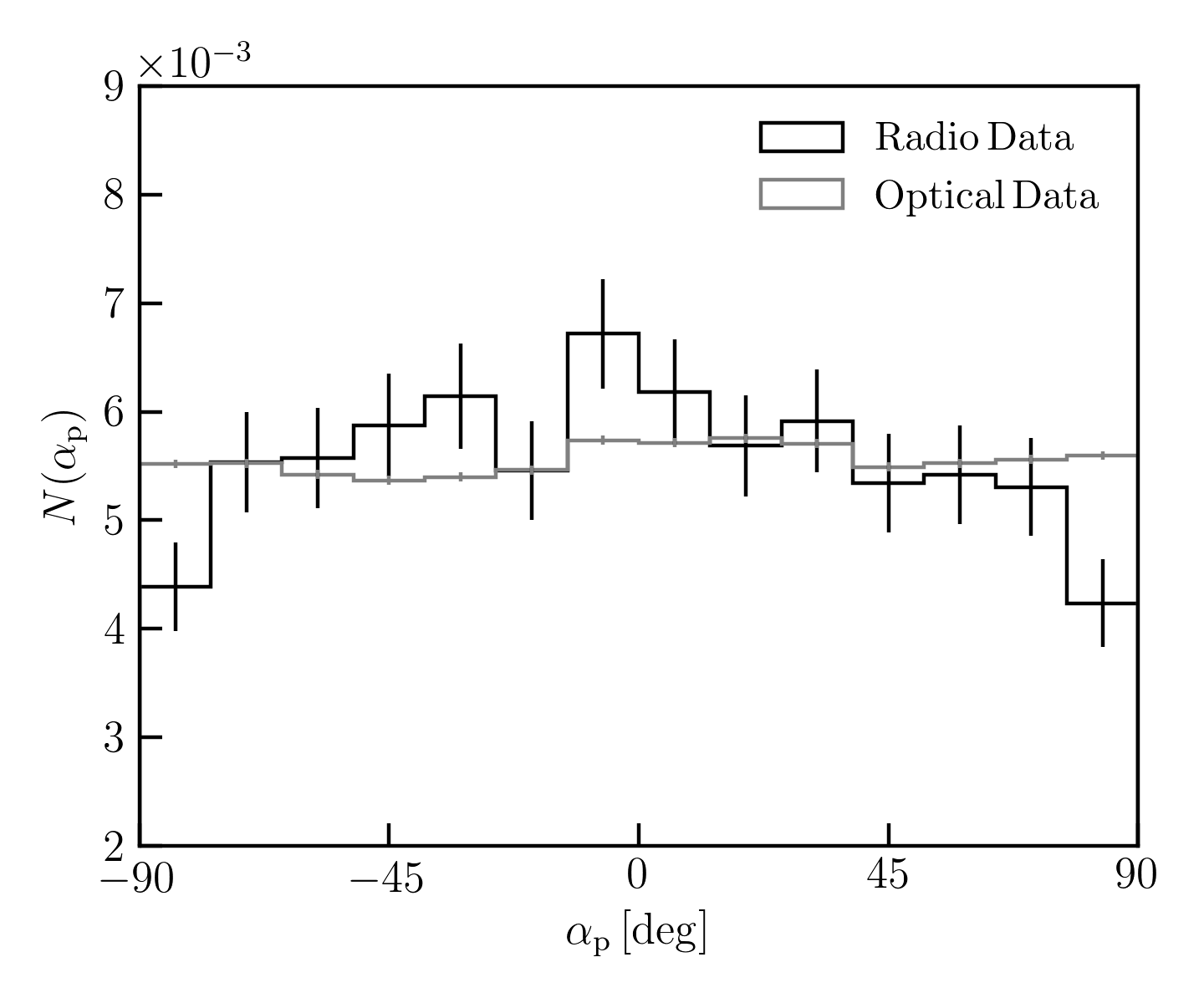}
    \caption{Normalised position angle distributions for the 3 GHz VLA radio and {\it HST}-ACS optical weak lensing catalogues. The vertical lines indicate Poisson counting errors.}
    \label{fig:hist_pos_ang_both}
\end{figure}

The position angle distributions are shown in Fig.~\ref{fig:hist_pos_ang_both}. The distribution for the optical data is consistent with uniform while there is a slight indication of a deviation from uniform in the radio position angle distribution. This could be indicative of a small unaccounted-for residual ellipticity in the PSF of the radio image (as described in Section~\ref{sec:radioshapes}, the radio shapes were extracted assuming a circular Gaussian PSF). However, any associated systematic effect present in the radio shape catalogue will be much smaller than the random noise, given the low number density of sources. Moreover, in comparison to a flat distribution, we measure a reduced chi-squared value of $\chi^2_{\rm red} = 2.1$, for the position angle distribution of the radio weak lensing catalogue, which we deem to be suitable for our purposes.

The spatial distributions of the two weak lensing catalogues are shown in Fig.~\ref{fig:ScatterMap_afterCuts} and Table~\ref{tab:WLcatalogues} lists the summary statistics for the two catalogues. The scatters, $\sigma_{\epsilon_{1,2}}$ in Table~\ref{tab:WLcatalogues} are the standard deviations of the shape components in the weak lensing catalogues, and give a measure of the intrinsic ellipticity dispersion in each catalogue, including measurement uncertainties.

\begin{table}
	\centering
	\caption{Summary statistics for the two weak lensing catalogues.}
	\label{tab:WLcatalogues}
	\begin{tabular}{lcccc} 
		\hline
		Survey & Total Number & Source & \multicolumn{2}{c}{Scatter} \\
		 & of Sources & Density & $\sigma_{\epsilon_1}$ & $\sigma_{\epsilon_2}$ \\
		 & & [arcmin$^{-2}$] & & \\
		\hline
		3 GHz VLA & 2,038 & 0.2 & 0.304 & 0.322 \\
		\textit{HST}-ACS & 243,852 & 41 & 0.270 & 0.267 \\
		\hline
	\end{tabular}
\end{table}

\section{Radio-Optical Shape Comparison}
\label{sec:shapecomparison}
\begin{figure}
	\includegraphics[width=\columnwidth]{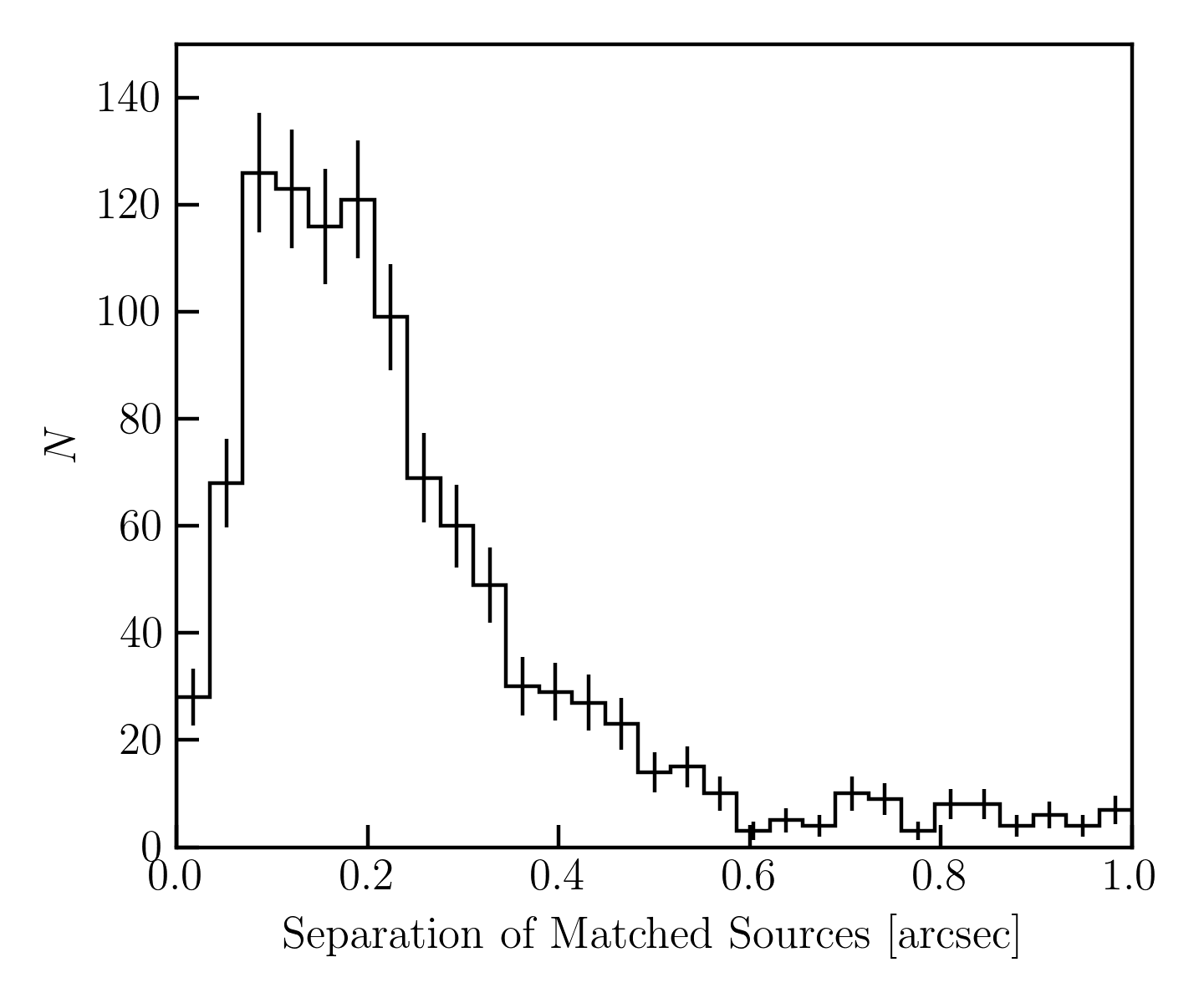}
    \caption{Separations of the 3 GHz VLA radio sources and their cross-matched {\it HST}-ACS optical counterparts, both from the weak lensing catalogues. We find 66\% (53\%) of the radio sources have an optical counterpart out to a maximum separation of 1$^{\prime \prime}$ (0.4$^{\prime \prime}$). The error bars show Poisson counting errors.}
    \label{fig:hist_separations}
\end{figure}

In order to compare the shapes of sources common to both samples, the 3 GHz VLA and {\it HST}-ACS weak lensing catalogues were first cross-matched, centred on the 3 GHz VLA sources. Fig.~\ref{fig:hist_separations} shows the separation of the resulting cross-matched sources. Sources that were matched to within 0.4$^{\prime \prime}$ were regarded as associated and used in the analysis. The percentage of radio sources remaining after this cut was 53\% (1,078 sources), equivalent to 0.44\% of sources in the optical catalogue.

\begin{figure}
	\includegraphics[width=\columnwidth]{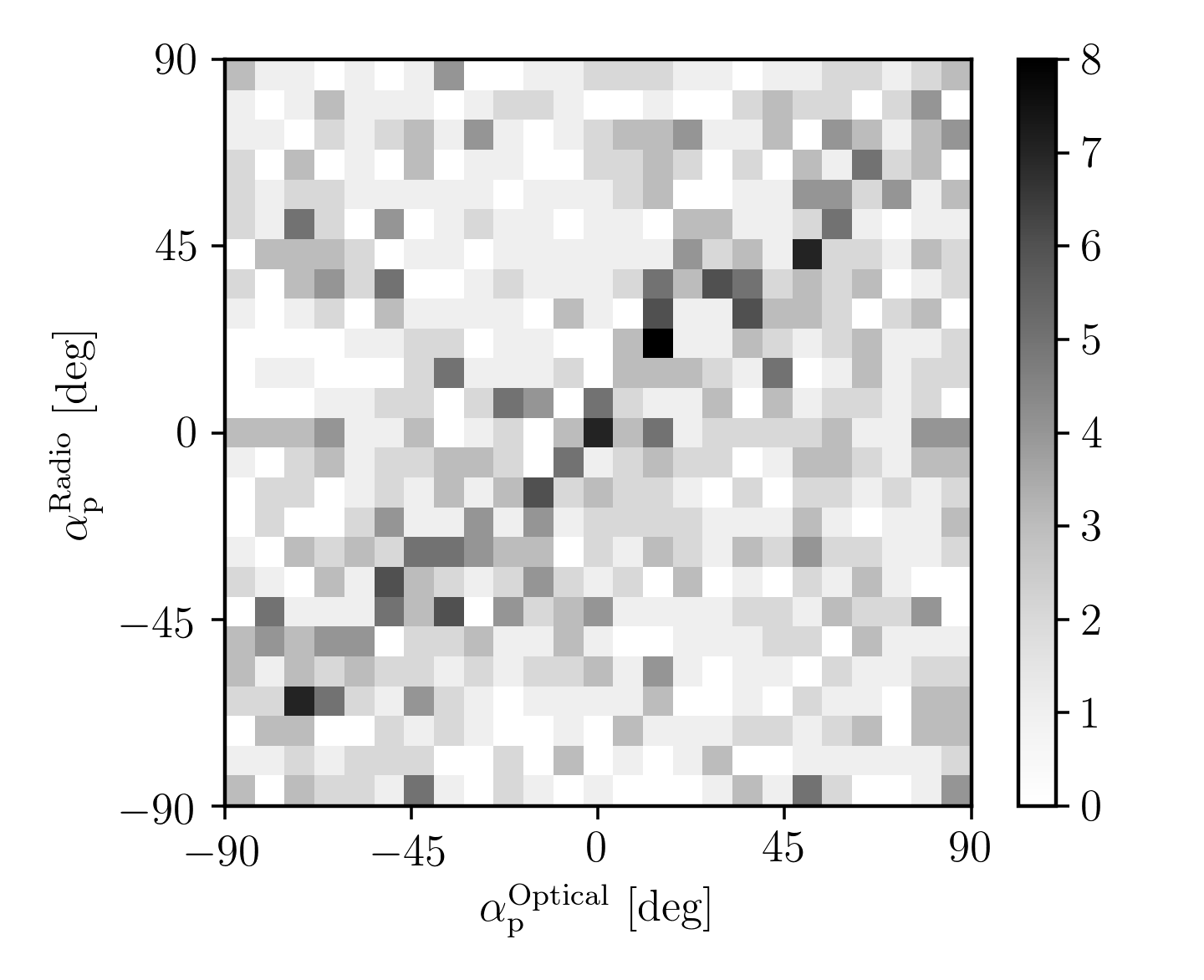}
    \caption{2D histogram of the position angle comparison between the matched 3 GHz VLA radio and {\it HST}-ACS optical weak lensing catalogues. The histogram contains 1,078 matched sources and the greyscale bar indicates the source number density.}
    \label{fig:alpha_Comparison_Op_3GHz_2dhist}
\end{figure}

\begin{figure}
	\subfigure{\includegraphics[width=\columnwidth]{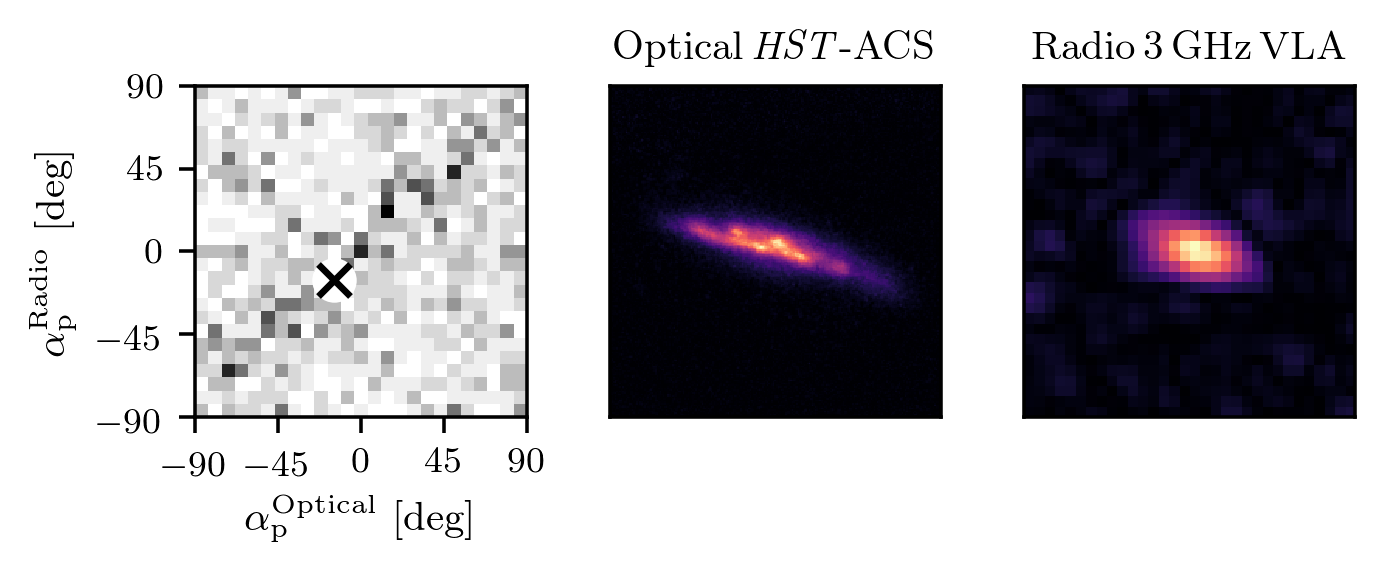}}\vspace{-0.25cm}%
	\vspace{-0.25cm}
    \subfigure{\includegraphics[width=\columnwidth]{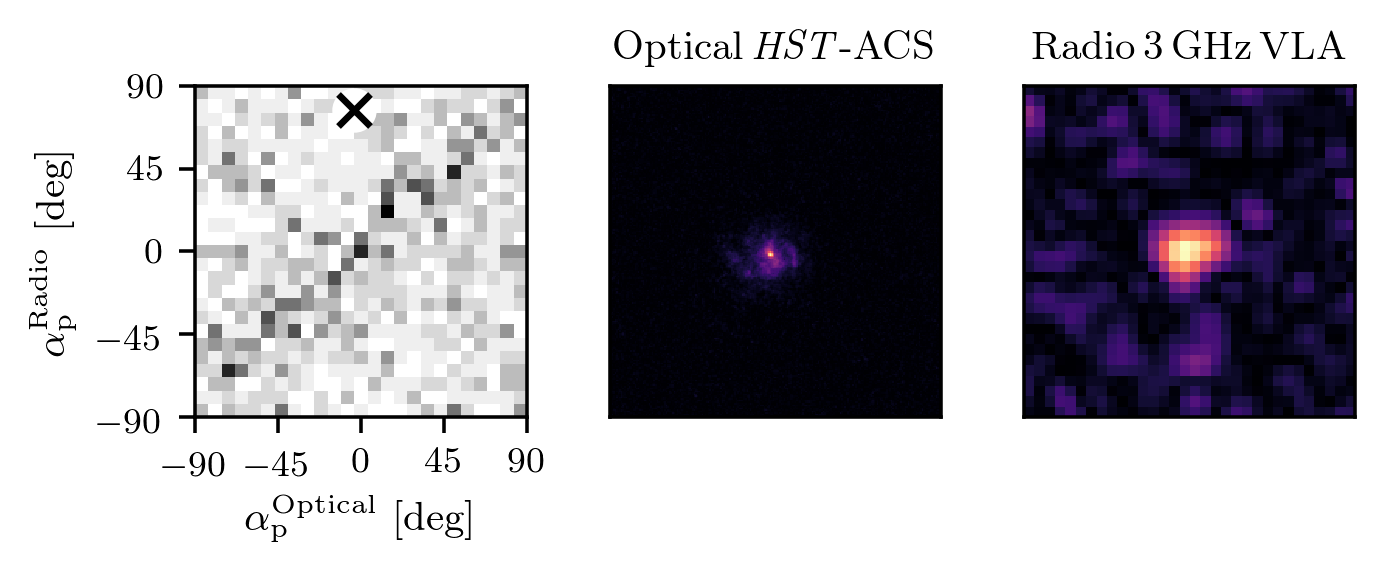}}\vspace{-0.25cm}%
	\vspace{-0.25cm}
    \subfigure{\includegraphics[width=\columnwidth]{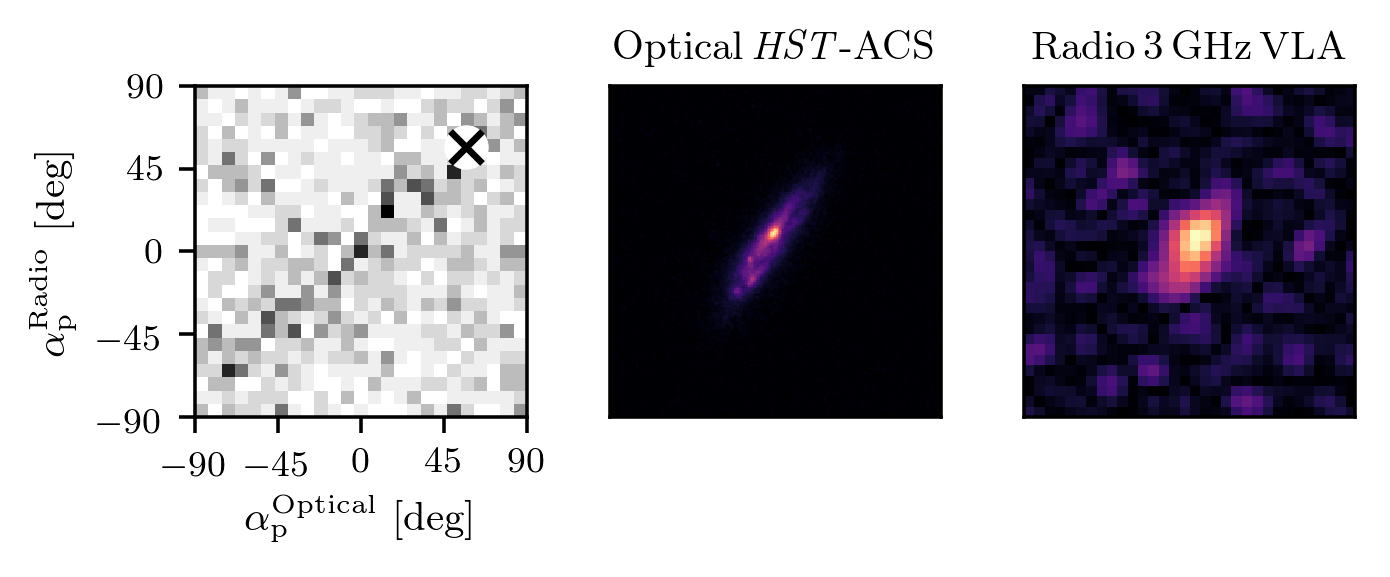}}\vspace{-0.25cm}%
	\vspace{-0.25cm}
    \subfigure{\includegraphics[width=\columnwidth]{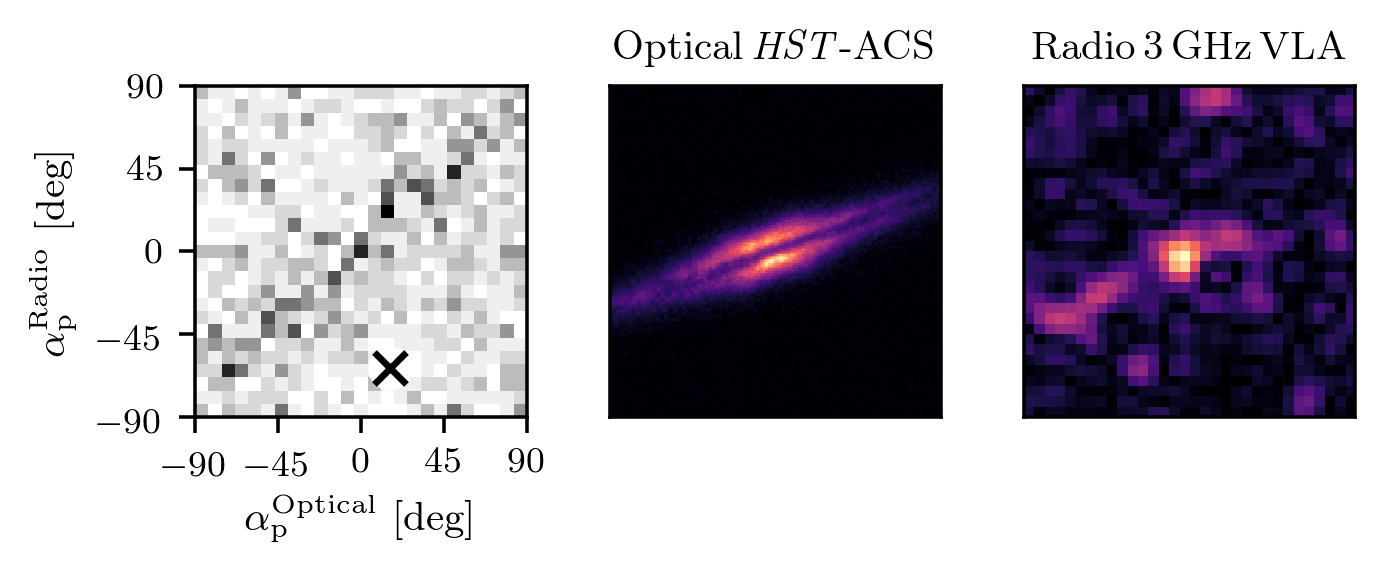}}\vspace{-0.25cm}%
	\vspace{-0.25cm}
    \subfigure{\includegraphics[width=\columnwidth]{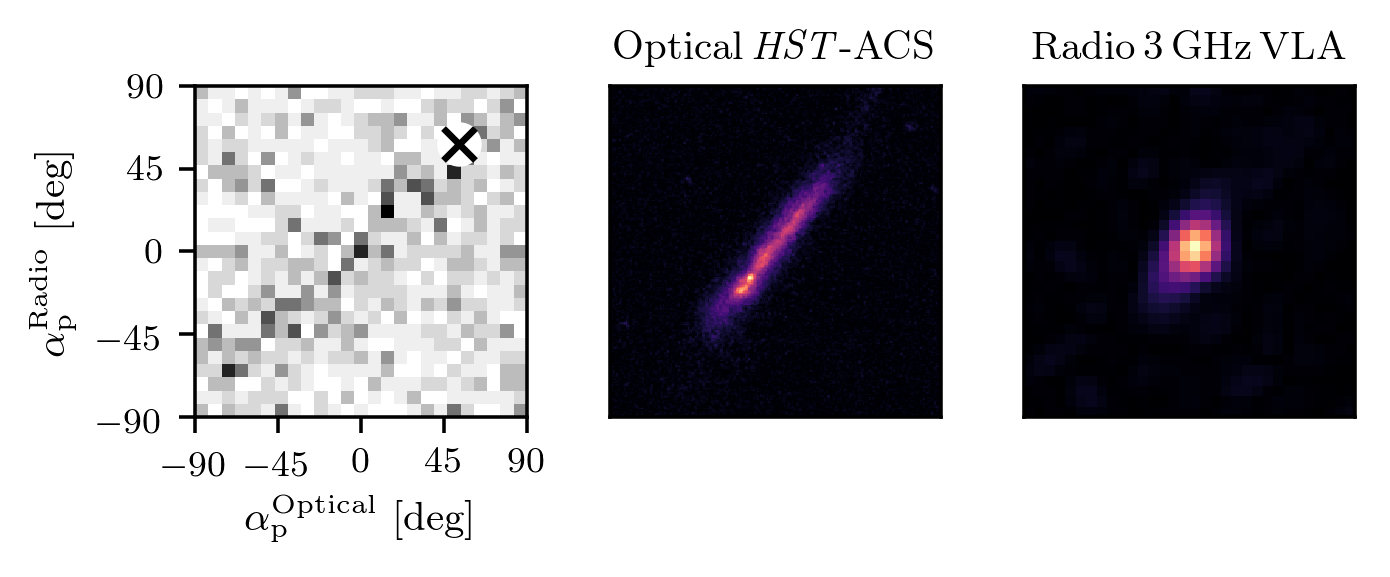}}\vspace{-0.25cm}%
	\vspace{-0.25cm}
    \subfigure{\includegraphics[width=\columnwidth]{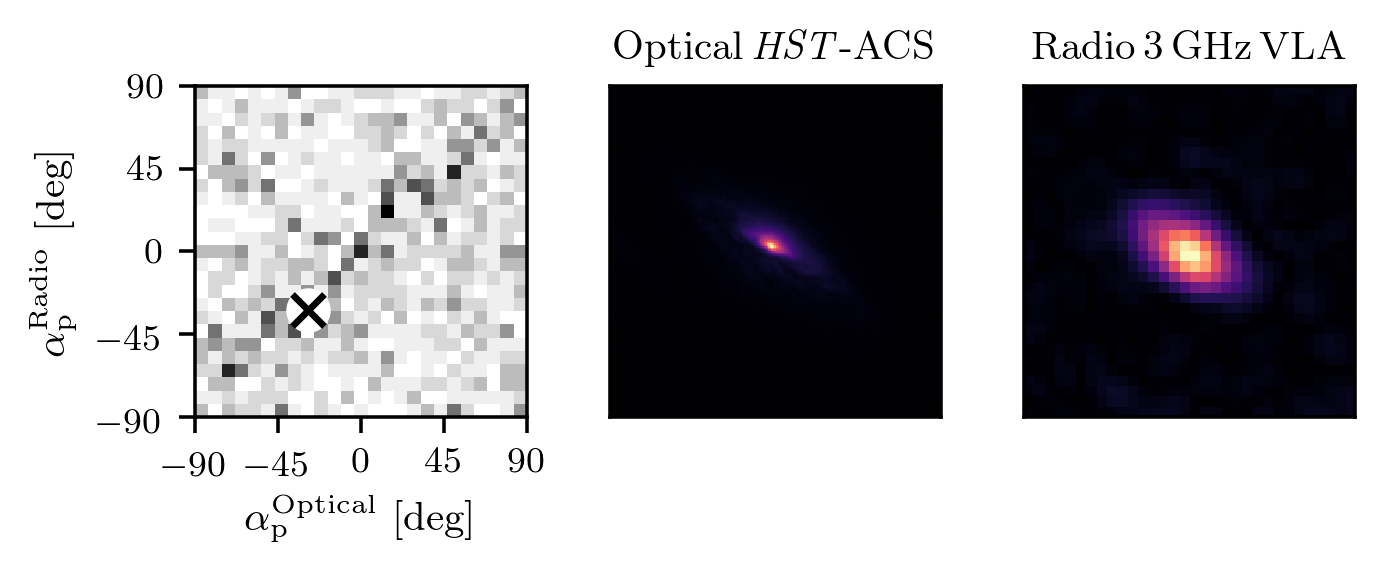}}
    \caption{Cut-out images for some of the cross-matched sources from the \textit{HST}-ACS and 3 GHz VLA catalogues. For each source (each row of plots), the position angle comparison histogram from Fig.~\ref{fig:alpha_Comparison_Op_3GHz_2dhist} is reproduced, with the measured correlation for the selected source indicated by the cross. For comparison, equivalent cut-outs for the lower resolution 1.4 GHz VLA-COSMOS survey can be seen in Fig.~10 of Tunbridge et al. (2016).}
    \label{fig:cutouts}
\end{figure}

To compare the radio and optical galaxy shapes, we compare the position angles, equation~(\ref{eq:posangle}), of the cross-matched radio and optical sources. Fig.~\ref{fig:alpha_Comparison_Op_3GHz_2dhist} shows a 2D histogram of this comparison. The figure clearly shows a significant correlation between the position angles of the galaxies in the two surveys. To quantify the strength of the observed correlation, we measure a Pearson correlation coefficient~\footnote{A perfect positive correlation would give $R_{\alpha} = 1$, and no correlation would give $R_{\alpha} = 0$. The uncertainty is given by $\sigma_{R_\alpha} = \sqrt{(1-R_\alpha^2)/(N-2)}$, where $N$ is the number of matched sources.} of $R_{\alpha} = 0.14 \pm 0.03$. This can be compared to the results from previous work, where Pearson correlation coefficient values of 0.028~\citep{Tunbridge2016} and $0.097 \pm 0.090$~\citep{Patel2010} were found. However, we urge caution when comparing the correlation coefficients measured in different analyses. Note, in particular, that the relation between the measured correlation coefficient and the degree of correlation is not, in general, linear. For example, while  \cite{Patel2010} measured a correlation coefficient $\sim0.1$ close to the value measured here ($0.14$), their analysis did not {\it detect} a correlation, due in part to the low number of galaxies included in their sample (123 compared to the 1,078 sources used here). In contrast, it is clear from Fig.~\ref{fig:alpha_Comparison_Op_3GHz_2dhist} that the radio and optical position angles of the sources examined here are significantly correlated with one another. In Fig.~\ref{fig:cutouts}, we present the radio and optical images for a selection of the cross-matched galaxies, alongside the corresponding position of those objects in the 2D histogram position angle comparison plot.

The study of this correlation gives insight into the intrinsic emission processes of optical and radio galaxies. However, it does not represent a detection of shear correlations. (Our shear correlation constraints are presented in Section~\ref{sec:powerspectra}, where we estimate the radio-optical cosmic shear cross-power spectrum.) Rather, the correlation that we present in Fig.~\ref{fig:alpha_Comparison_Op_3GHz_2dhist} is dominated  by the intrinsic shapes of the galaxies. The detected correlation suggests that the orientations of galaxies are unchanged whether they are viewed in the optical or radio bands, which agrees with the expectation that the source of the synchrotron emission traced by the radio data is co-located with the stars responsible for the optical emission.

\section{Cosmic Shear Power Spectra}
\label{sec:powerspectra}
In this section, we estimate the three possible cosmic shear power spectra ($C_\ell^{\rm OO}, C_\ell^{\rm RO}$ and $C_\ell^{\rm RR}$) from the optical (O) and radio (R) shear catalogues. To do this, we use a flat-sky maximum likelihood power spectrum estimation code~\footnote{Publicly available from \url{https://bitbucket.org/fkoehlin/qe\_public}}, described fully in~\cite{Kohlinger2016, Kohlinger2017}. The code is based on the algorithm proposed in~\cite{Hu2001}, which was first applied to real data in the cosmic shear analysis of the COMBO-17 survey~\citep{Brown2003a}. The same method was also used to measure the shear power spectrum in Stripe 82 of the SDSS survey~\citep{Lin2012}. \cite{Kohlinger2016} adapted the technique to enable a tomographic analysis, which they applied to the CFHTLenS data. More recently the technique has been used to perform a tomographic shear analysis of the KiDS-450 data~\citep{Kohlinger2017}. This analysis found results consistent with the real-space analysis of \cite{Hildebrandt2017} when appropriate scale cuts were used, in particular when small angular scales were excluded. We do not apply such cuts in this work because the COSMOS field only spans $2 \deg^2$. Hence, there is a large overlap of the scales probed here and the previous real-space studies~\citep{Massey2007, Schrabback2010} and so the discrepancy between real-space and power spectra measurements is not expected to be as large as in the KiDS-450 studies.

We have chosen to use this particular code because of its ability to extract the cross-power spectra between two shear maps with uncorrelated shape noise (in addition to the auto-spectra). In the case of KiDS-450, this was applied to several pairs of redshift bins to produce a tomographic analysis. However, the same technique can also be used to extract the cross-spectra between different wavebands, in our case the optical and radio bands. The only extension which may be required for future cross-waveband analyses would be to account for the correlated shape noise term of equation~(\ref{eq:noisecrossfull}). In the case where there are significant numbers of matched sources between the optical and radio catalogues, and these sources have correlated intrinsic shapes in the radio and optical bands, this noise term will be non-zero and should be accounted for. However, we find that this modification is not necessary for the current analysis because of the very low number of matched sources compared to the number of sources in the optical catalogue. We discuss this further in Section~\ref{sec:crossnoiseconsiderations}. 

Note that we do not apply a tomographic shear analysis here. Instead we perform a 2D cosmic shear analysis, where all of the background sources within each shear catalogue are considered to reside in a single, very broad, redshift bin. We proceed initially by constructing pixelised shear maps from the two weak lensing shape catalogues described in Section~\ref{sec:sourceselection}. To construct these, we simply average the galaxy ellipticity component measurements ($\epsilon_1$ \& $\epsilon_2$) in square pixels of side length 3.4 arcmin, using uniform weights. Note that when assigning galaxies to shear pixels, a gnomonic projection is used to convert the WCS coordinates (RA, Dec) of each source to flat sky Cartesian coordinates. For the $2 \deg^2$ COSMOS field, this is an excellent approximation, following the discussion presented in Section~\ref{sec:coord_projections}. The average number of galaxies contributing to each shear pixel was 473 for the optical data and 3.5 for the radio data. The power spectrum extraction also requires an estimate of the intrinsic ellipticity dispersion in each sample, for which we used the values listed in Table~\ref{tab:WLcatalogues}.

\subsection{Band Power Selection}
\label{sec:bandpowerselection}

We estimate the various shear power spectra in ``band powers", i.e. the code estimates a weighted average of the quantity, $\ell(\ell+1)C_\ell / 2\pi$, across some nominal multipole range, $\Delta\ell$. Note, however, that the actual sensitivity of each band power, $P_b$, to a model cosmic shear signal, $C_\ell$, is more accurately described by the band power window functions, $W_{b\ell}$. These are defined such that the theoretical expectation value for each band power is given by~\citep[e.g.][]{Knox1999, Lin2012}
\begin{equation}
P_b^{\rm th}  = \sum_\ell \frac{W_{b\ell}}{\ell} \frac{\ell(\ell+1)}{2\pi} C_\ell.
\label{eq:band_expectation}
\end{equation}
The power spectrum code also calculates the $W_{b\ell}$ functions, according to the prescription detailed in \cite{Lin2012}.

For all three pairs of shear maps (OO, RO, and RR), in addition to the $E$-mode spectra ($C_\ell^{EE}$, which contain the cosmological signal), we also estimate the $B$-mode spectra, $C^{BB}_\ell$, and the cross-correlation between $E$- and $B$-modes, $C_\ell^{EB}$. As discussed in Section~\ref{sec:theory}, these latter two spectra are expected to be consistent with zero in the absence of instrumental or astrophysical systematics.

To select the nominal multipole ranges for power spectrum extraction, we follow the prescription given in~\cite{Kohlinger2017}. The largest multipole that can be extracted from the shear maps is determined by the shear pixel side length, $\theta_{\rm pix} = 3.4$ arcmin, which corresponds to $\ell_\mathrm{pix} = 6,360$. The smallest accessible multipole (largest angular scale) is determined by the survey area. For our analysis, we set the lowest multipole according to the survey side length of the \textit{HST}-ACS optical map, $\theta_\mathrm{max} = 1.28 \deg$, corresponding to $\ell_\mathrm{field} = 282$. When extracting the power spectra, we additionally include a ``junk" band power at lower multipoles (with $\ell_\mathrm{min} = 100$) that is intended to be sensitive to DC offset effects and/or ambiguous modes (modes which cannot be separated into $E$- and $B$-modes) only.  

The widths of the band powers were set to at least $2 \ell_\mathrm{field} \approx 570$, so that correlations between bands were minimised~\citep{Hu2001}. Finally, the maximum $\ell$ of the highest-$\ell$ band power (which was also included as a ``junk" band) was extended to $2 \ell_\mathrm{pix} \approx 12,800$, in order to absorb any effects from the highly oscillatory behaviour of the pixel window function on scales close to and larger than $\ell_\mathrm{pix}$~\citep{Kohlinger2017}.

In addition to the first and last band powers (which were discarded for the reasons described above), we have followed \cite{Kohlinger2017} in also discarding the second-to-last band power when interpreting the results. However, the recovered errors on this band power are large (for all of our power spectra channels) and we have checked that   its inclusion does not significantly change our conclusions.

The resulting band power definitions are listed in Table~\ref{tab:bandpowerselection}. The table also lists the nominal $\theta$-ranges for each bin. However these serve only as an approximate indication of the real-space scales probed by each band power. They should not be used to directly compare the power spectrum measurements to real-space correlation function analyses~\citep[see][for further discussion]{Kohlinger2017}.

\begin{figure}
	\includegraphics[width=\columnwidth]{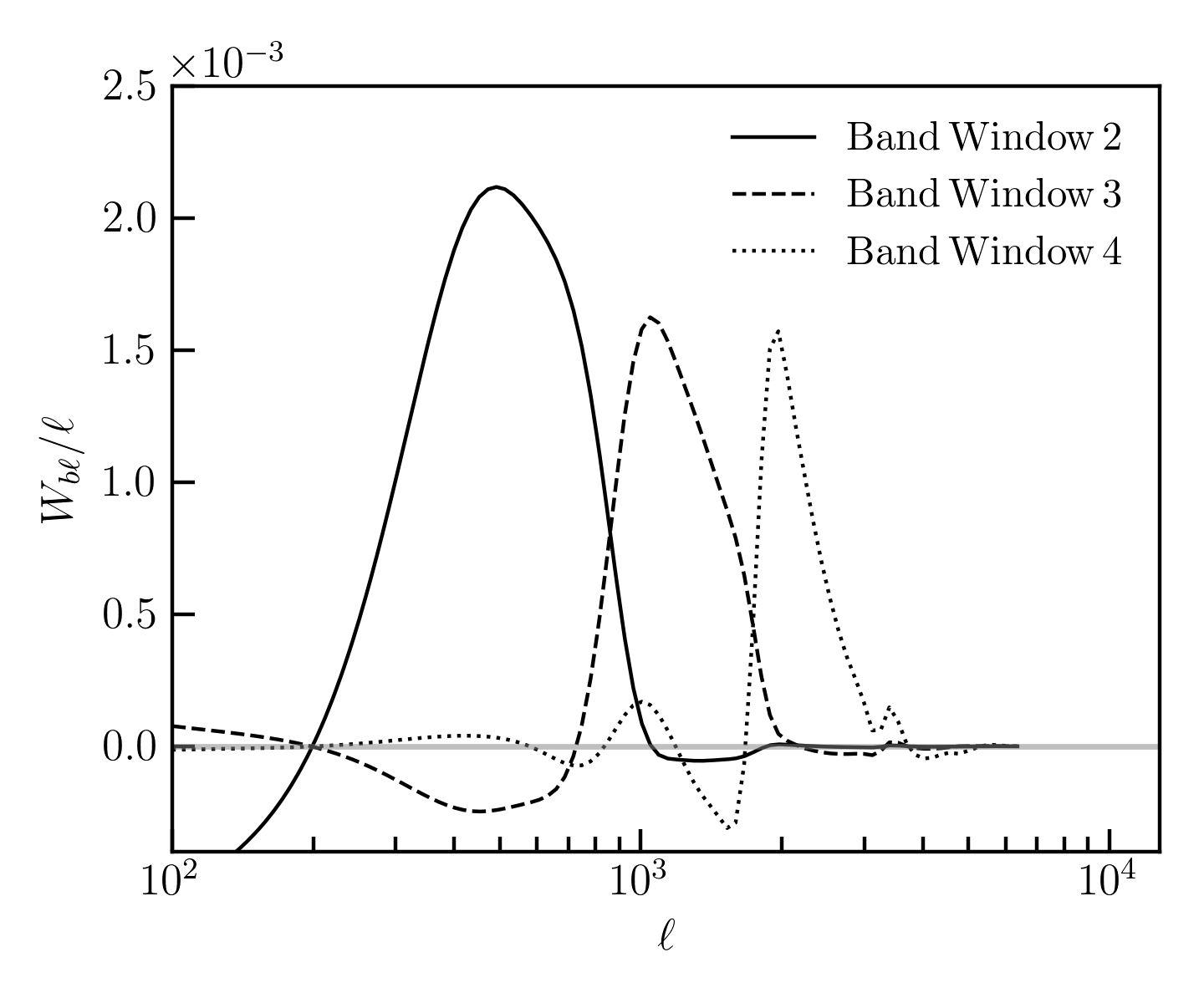}
    \caption{The band power window functions for bins 2, 3 and 4, for the $E$-mode optical auto-power spectra measurements. The window functions for the $E$-mode radio-optical cross- and radio auto-power spectra are broadly similar.}
    \label{fig:bandwindowfunctions}
\end{figure}

\begin{figure}
	\includegraphics[width=\columnwidth]{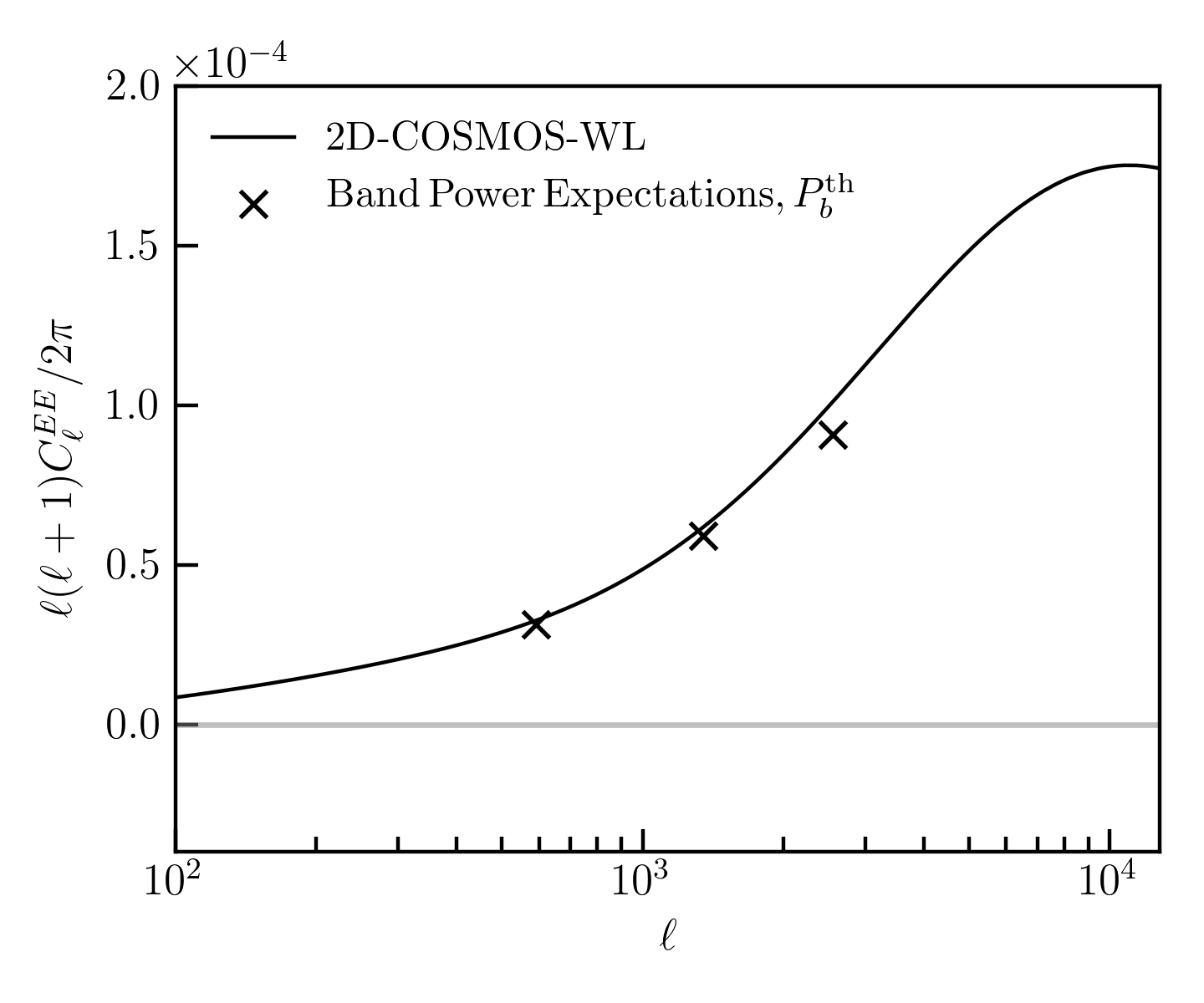}
    \caption{The expectation values for band powers 2, 3 and 4, for a specific cosmological model (the 2D-COSMOS-WL best-fit model for the optical-optical auto-power spectrum with $z^{\rm O}_{\rm{med}} = 1.26$ described in Section~\ref{sec:GRFclones}). These have been calculated by convolving the band power window functions shown in Fig.~\ref{fig:bandwindowfunctions} with the 2D-COSMOS-WL theory power spectrum (plotted here as the smooth curve), as in equation~(\ref{eq:band_expectation}). The band power expectations for the radio-optical cross- and radio auto-power spectra follow a similar form.}
    \label{fig:bandexpectations}
\end{figure}

In Fig.~\ref{fig:bandwindowfunctions}, we plot the band power window functions, $W_{b\ell}$, for the three $C_\ell^{\rm OO}$ ($E$-mode) band powers that are retained for the science analysis. (The window functions for the $C_\ell^{\rm RO}$ and $C_\ell^{\rm RR}$ spectra are very similar.) The negative side lobes of these functions are a well-known feature of maximum-likelihood $C_\ell$-estimators and result in neighbouring band powers being slightly anti-correlated. The expectation values (equation~\ref{eq:band_expectation}) of the same three band powers for an example cosmological model are displayed in Fig.~\ref{fig:bandexpectations}, alongside the corresponding model shear power spectrum. This figure shows that the expectation values closely trace the underlying power at the central multipole of each band, and justifies the use of a simple visual comparison between the band power measurements and theoretical shear power spectra for specific cosmological models.

\begin{table}
	\centering
	\caption{Band power intervals used for the power spectra extraction. Only bins 2, 3 and 4 were retained for further analysis, see Section~\ref{sec:bandpowerselection}.}
	\label{tab:bandpowerselection}
	\begin{tabular}{ccc} 
		\hline
		Band No. & $\ell$-range & $\theta$-range\\
         & & [arcmin] \\
		\hline
		\color{mygrey}{1} & \color{mygrey}{100-281} & \color{mygrey}{216.0-76.9} \\
		2 & 282-899 & 76.6-24.0 \\
		3 & 900-1799 & 24.0-12.0 \\
        4 & 1800-3299 & 12.0-6.5 \\
        \color{mygrey}{5} & \color{mygrey}{3300-5999} & \color{mygrey}{6.5-3.6} \\
        \color{mygrey}{6} & \color{mygrey}{6000-12800} & \color{mygrey}{3.6-1.7} \\
		\hline
	\end{tabular}
\end{table}

\begin{figure*}
	\includegraphics[width=\textwidth]{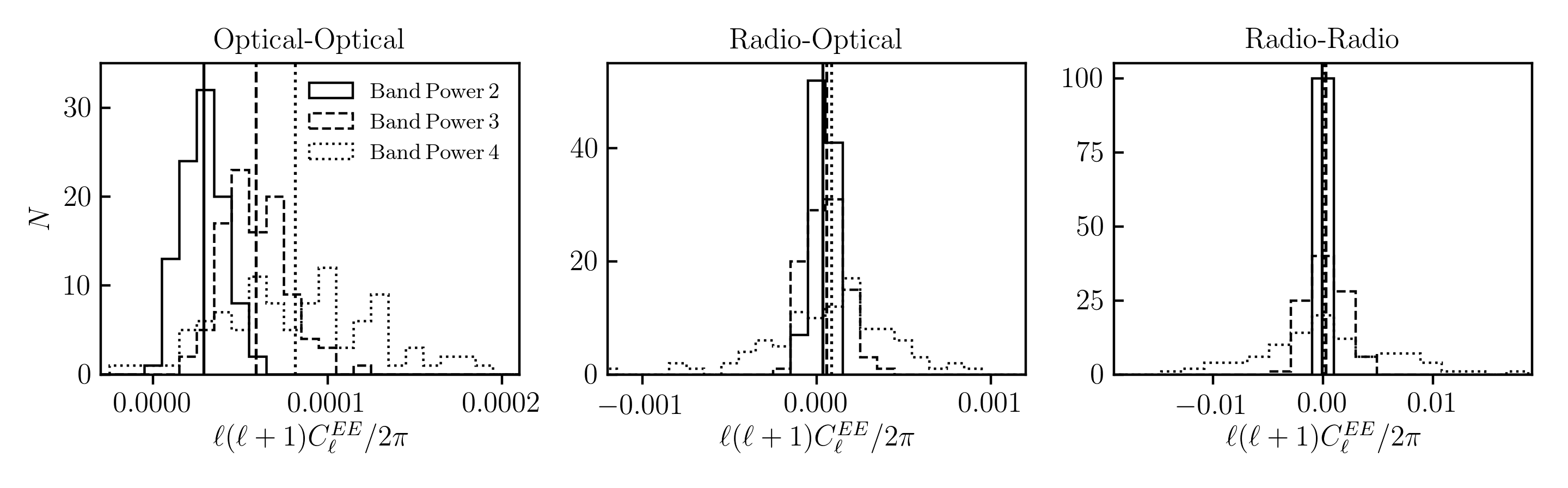}
    \caption{Distributions of the $C_\ell^{EE}$ band powers recovered from the 100 mock datasets. The left panel shows the $C_\ell^{\rm OO}$ band powers, the centre panel shows the $C_\ell^{\rm RO}$ band powers and the right panel shows the $C_\ell^{\rm RR}$ band powers. The vertical lines show the mean recovered band powers. Note the different vertical and horizontal scales for all three plots. The $C_\ell^{BB}$ and $C_\ell^{EB}$ band power distributions look broadly similar.}
    \label{fig:GRF_bandpowers_hist}
\end{figure*}

\begin{figure*}
	\includegraphics[width=\textwidth]{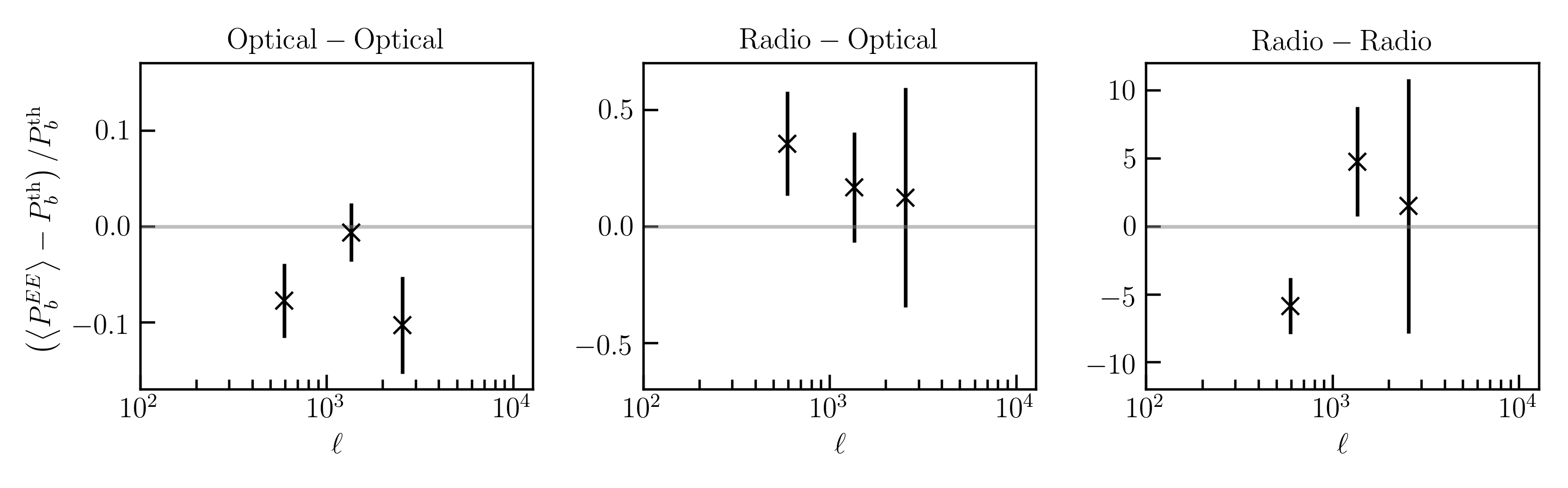}
    \caption{Residuals between the mean $E$-mode band powers recovered from the 100 mock datasets, $\left < P_b^{EE} \right >$ (vertical lines in Fig.~\ref{fig:GRF_bandpowers_hist}) and the expectation band powers, $P_b^{\rm th}$ calculated using equation~(\ref{eq:band_expectation}). The three panels represent the same sets of measurements as Fig.~\ref{fig:GRF_bandpowers_hist}. The vertical error bars depict 1$\sigma$ errors on the mean values, calculated by scaling the standard deviations of the band powers measured from the $N=100$ mock datasets by $1/\sqrt{N}$. Note the different vertical scales for all three plots. The left-hand panel can be compared to Fig.~1 of K\"{o}hlinger et al. (2016) and Fig.~D1 of K\"{o}hlinger et al. (2017).}
    \label{fig:GRF_bandpowers_diffs}
\end{figure*}

\subsection{Simulations}
\label{sec:GRFclones}
To estimate the uncertainties in the band power measurements, we use Monte Carlo simulations of the optical and radio shape catalogues. We also use these simulations to assess the performance of the power spectrum extraction in the presence of the masking and spatial distribution of sources found in the real data. The simulated catalogues are based on mock shear fields, which we generate as Gaussian Random Fields (GRFs). We generate the input shear power spectra according to equation~(\ref{eq:powspecfull}). For the background cosmology, we adopt a flat, 6-parameter $\Lambda$CDM model, assuming the best-fitting parameter values derived from the 2D weak lensing analysis of the \textit{HST}-COSMOS data by~\cite{Schrabback2010} (hereafter referred to as 2D-COSMOS-WL). (We note that the \cite{Schrabback2010} analysis fixed the values of $\Omega_\mathrm{b}$, $h$ and $n_\mathrm{s}$.) We have used the 2D-COSMOS-WL values since they were estimated using the same \textit{HST}-ACS $1.64 \deg^2$ COSMOS observations that we use here. One would therefore expect simulations based on the 2D-COSMOS-WL values to closely mimic the cosmic shear signal present in our weak lensing shape catalogues. The parameter values that we have used for generating the simulations are listed in Table~\ref{tab:cosmologymodel}. For comparison, the table also includes the best-fitting ``TT,TE,EE+lowP+lensing" analysis values from \cite{Planck2016}.

When generating theory power spectra for these models, we do not include the effects of intrinsic alignments ~\citep[e.g.][]{Heavens2000, Catelan2001, Crittenden2001, Brown2002, Troxel2015} or baryonic feedback effects \citep[e.g.][]{vanDaalen2011} on the observable shear power spectra since they are expected to contribute small corrections, which would be sub-dominant to our statistical error bars for sources in redshift regimes probed by the optical sources. For the higher redshift range probed by the radio sources, this is uncertain and may no longer be the case.

For the redshift distribution of the background sources, we use the parameterised form, 
\begin{equation}
    P(z) \propto z^2 \exp \left[ - \left( \frac{1.41 z}{z_{\rm{med}}} \right)^{1.5} \right].
	\label{eq:redshiftdist}
\end{equation}
For the median redshift of the optical sample, we use $z^{\rm O}_{\rm{med}} = 1.26$, as estimated by \cite{Massey2007} for the \textit{HST}-COSMOS dataset. The median redshift for the radio sample is less well known. Here we use a value of $z^{\rm R}_{\rm{med}} = 1.00$, which is consistent with expectations based on the source flux distribution in the 3 GHz COSMOS data, the SKA Design Study (SKADS) simulations of \cite{Wilman2008} and the updated T-RECS simulation of \cite{Bonaldi2018}.

\begin{table}
	\centering
	\caption{Background cosmology parameter values. The simulated datasets were generated using the 2D-COSMOS-WL parameter values, see Section~\ref{sec:GRFclones}. The {\it Planck} model values are from the 2015 {\it Planck} ``TT,TE,EE+lowP+lensing" analysis, and are included for comparison only.}
	\label{tab:cosmologymodel}
	\begin{tabular}{c c c} 
		\hline
        Parameter & 2D-COSMOS-WL & {\it Planck} 2015 \\
        \hline
        $\sigma_\mathrm{8}$ & 0.68 & 0.8150 \\
        $\Omega_{\mathrm{m}}$ & 0.30 & 0.3121 \\
        $\Omega_{\Lambda}$ & 0.70 & 0.6879 \\
        $\Omega_\mathrm{b}$ & 0.044 & 0.049 \\
        $h$ & 0.72 & 0.6727 \\
        $n_\mathrm{s}$ & 0.96 & 0.9653 \\
        \hline
	\end{tabular}
\end{table}

Once generated, we sample the simulated GRF shear fields using the real source positions as listed in the optical and radio shape catalogues, described in Section~\ref{sec:sourceselection}. This process results in pairs of mock optical and radio shear catalogues with identical entries to the real catalogues except that the real data ellipticity measurements are replaced with the appropriate shear values from the GRF simulations. We add noise to each mock ellipticity measurement by randomly selecting a measured ellipticity from the real data shape catalogue and adding it to the simulated ellipticity entry. Note that this process models both the intrinsic shape noise and measurement error contributions to the measured ellipticities such that the ellipticity distributions in the mock catalogues closely follow those of the real data (Fig.~\ref{fig:hist_ellipticity_both}). This simulation technique also naturally replicates the spatial distribution of sources in the real data (Fig.~\ref{fig:ScatterMap_afterCuts}), including for example, the gaps in the coverage of the optical data due to masking. 

A total of 100 different pairs of simulated catalogues were generated following this procedure. The resulting catalogues were analysed with the power spectrum estimator in exactly the same way as was used for the analysis of the real data. Error bars were calculated as the standard deviations of the 100 recovered band powers for each multipole bin. In doing this, we assume a symmetrical distribution of band powers around the mean for each bin (not necessarily Gaussian) which we find to be a good assumption upon plotting histograms of the output band powers. This is demonstrated in Fig.~\ref{fig:GRF_bandpowers_hist} which shows the distributions of the $C_\ell^{EE}$ band power values recovered from the simulations. The equivalent distributions for the $C_\ell^{BB}$ and $C_\ell^{EB}$ band powers look broadly similar.

The accuracy with which we recover the input shear power spectra is illustrated in Fig.~\ref{fig:GRF_bandpowers_diffs}, which shows the residuals between the mean $E$-mode band powers recovered from the 100 mock datasets, $\left < P_b^{EE} \right >$ and the expectation band powers, $P_b^{\rm th}$ calculated using equation~(\ref{eq:band_expectation}). The expectation band powers for the optical-optical spectrum were shown in Fig.~\ref{fig:bandexpectations}. We recover the input spectra to a sufficient level of accuracy and precision for all three channels with reduced chi-squared values of 2.7, 1.0 and 3.1 for the optical-optical, radio-optical and radio-radio $E$-mode band powers respectively.

\begin{figure*}
	\includegraphics[width=\textwidth]{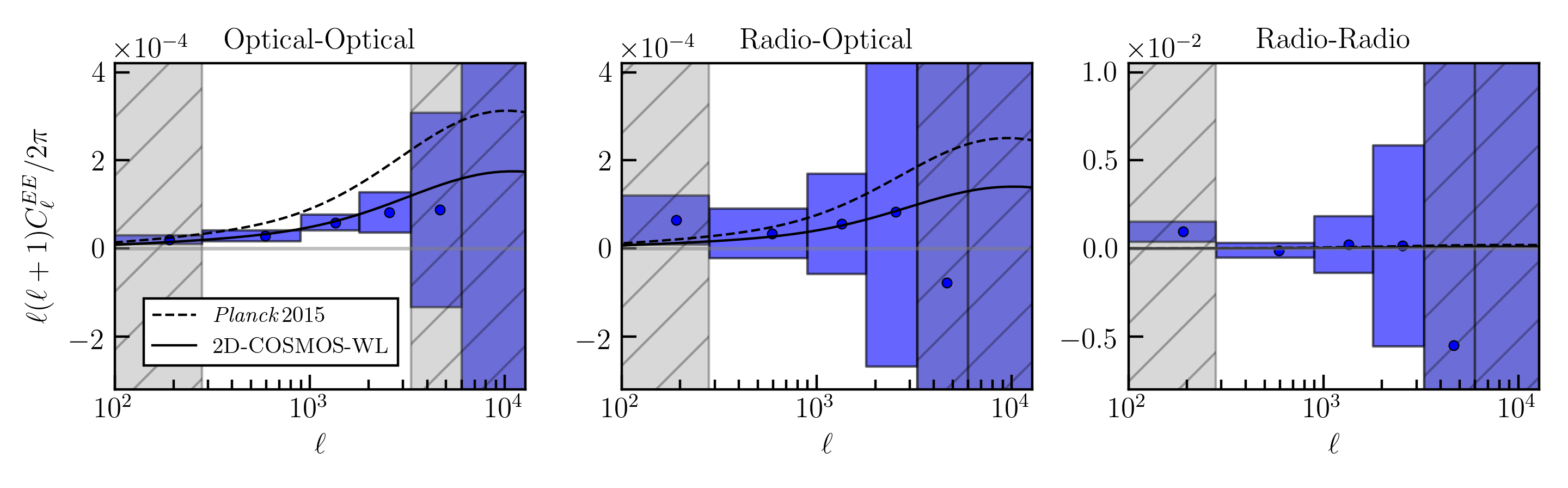}
    \caption{The mean $E$-mode band powers recovered from the 100 mock datasets. From left to right: optical-optical, radio-optical and radio-radio. The points show the mean recovered band powers for each multipole bin, and the error boxes indicate the standard deviations of the band powers. Note the different vertical axis scale of the radio-radio power spectrum to accommodate the larger error bars. The two model spectra curves correspond to the two cosmological models summarized in Table~\ref{tab:cosmologymodel} and make use of the full redshift distribution given in equation~(\ref{eq:redshiftdist}) with redshift combinations: $z_{\rm med}^{\rm O} \times z_{\rm med}^{\rm O}$, $z_{\rm med}^{\rm R} \times z_{\rm med}^{\rm O}$ and $z_{\rm med}^{\rm R} \times z_{\rm med}^{\rm R}$ from left to right, where $z_{\rm med}^{\rm O} = 1.26$ and $z_{\rm med}^{\rm R} = 1.00$.}
    \label{fig:GRF_powerSpectra_EE}
\end{figure*}

\begin{figure*}
	\includegraphics[width=\textwidth]{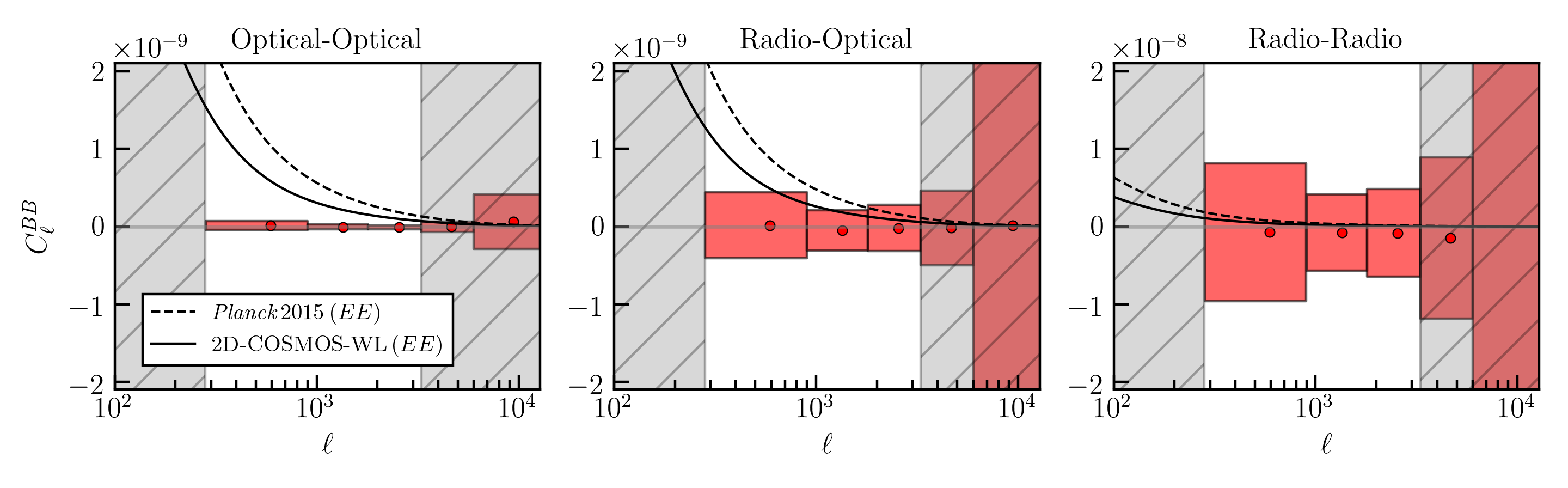}
    \caption{Same as Fig.~\ref{fig:GRF_powerSpectra_EE} but for the $B$-mode power spectra recovered from the 100 mock datasets. In all cases, the recovered $B$-mode power spectra are consistent with zero. Note the different band power and vertical axis scaling in comparison to Fig.~\ref{fig:GRF_powerSpectra_EE} as well as the different vertical axis scale of the radio-radio spectrum. The models shown are the same theoretical $E$-mode spectra as shown in Fig.~\ref{fig:GRF_powerSpectra_EE}.}
    \label{fig:GRF_powerSpectra_BB}
\end{figure*}

\begin{figure*}
	\includegraphics[width=\textwidth]{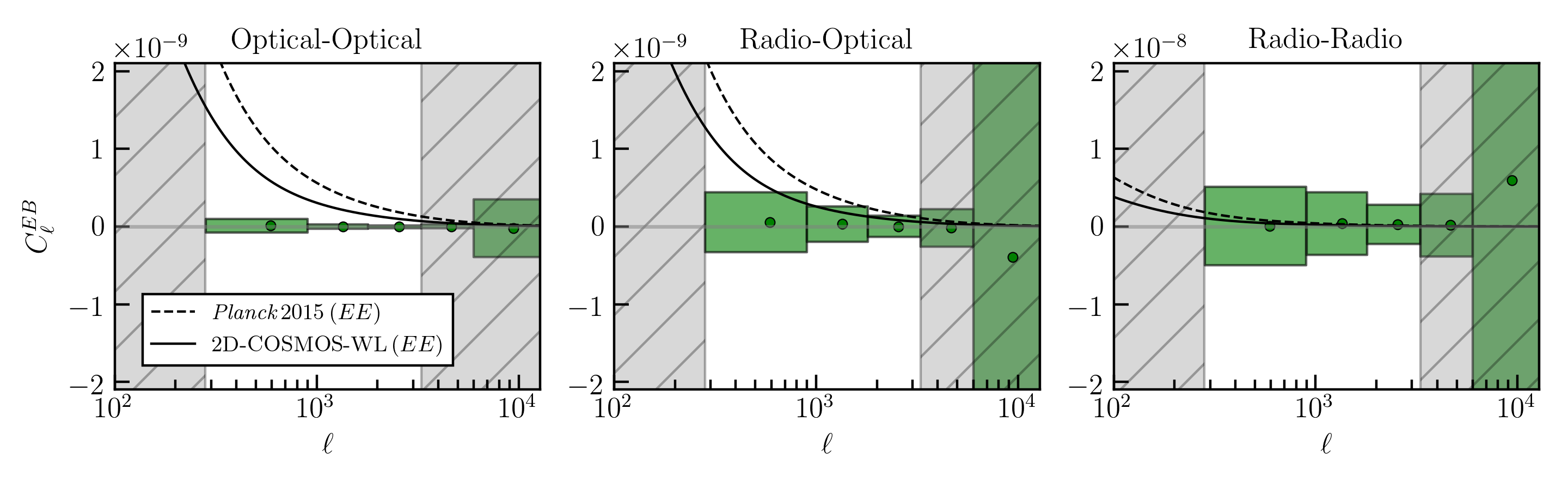}
    \caption{Same as Fig.~\ref{fig:GRF_powerSpectra_BB} but for the $EB$ cross-correlation power spectra extracted from the 100 mock datasets. Again, each of the recovered spectra are consistent with zero.}
    \label{fig:GRF_powerSpectra_EB}
\end{figure*}

The mean recovered band powers from the 100 mock datasets are shown in Fig.~\ref{fig:GRF_powerSpectra_EE} for the $C^{EE}_\ell$ spectra, in Fig.~\ref{fig:GRF_powerSpectra_BB} for the $C^{BB}_\ell$ spectra and in Fig.~\ref{fig:GRF_powerSpectra_EB} for the $C^{EB}_\ell$ spectra. These figures demonstrate that the input $E$-mode power spectra are recovered accurately by the maximum likelihood estimator with any residual biases being much smaller than the random errors. The recovered $B$-mode and $EB$ cross-correlation spectra are both consistent with zero, as expected. We note that the error bars for the optical-optical $E$-mode power spectrum ($C_\ell^{\rm OO}$) are dominated by cosmic variance (for the retained three band powers), whereas the error bars for the $E$-mode radio-optical cross-power ($C_\ell^{\rm RO}$) and for the $E$-mode radio-radio auto-power ($C_\ell^{\rm RR}$) are dominated by shape noise, a consequence of the low source number density in the radio catalogue. 

\subsection{Shear Measurement Calibration}
\label{sec:app_shear_calibration}
To validate our shear measurements, we measure calibration biases using simulated datasets with known shears. We do not calibrate our radio shear measurements primarily because a significant shear detection is not expected from either the $E$-mode radio-optical cross-power ($C_\ell^{\rm RO}$) or the $E$-mode radio-radio auto-power ($C_\ell^{\rm RR}$) spectra due to the limited precision available for those spectra, as shown in Fig.~\ref{fig:GRF_powerSpectra_EE}. Furthermore, accurate calibration of shear measurements from radio observations requires careful modelling of correlated noise and selection effects in the data, which we defer to a future study.

We extract calibration biases for our optical shear measurements using the commonly parameterised form,
\begin{equation}
	\gamma_i^{\rm meas} = (1 + m_i) \gamma_i^{\rm sim} + c_i
	\label{eq:shear_calibration},
\end{equation}
where $m_i$ and $c_i$ are the multiplicative and additive biases for the two shape components $i = \{1,2\}$. $\gamma_i^{\rm sim}$ denote input, simulated shear values, and $\gamma_i^{\rm meas}$ are shear values extracted via the measurement process of~\cite{Tunbridge2016}. For the case where additive biases are negligible, the multiplicative biases would be used to calibrate the shear power spectrum according to
\begin{equation}
	C_\ell^{\rm{cal}} = \frac{C_\ell^{\rm{uncal}}}{(1+m_1) (1+m_2)}
	\label{eq:shear_calibration_Cl},
\end{equation}
where $C_\ell^{\rm{cal}}$ and $C_\ell^{\rm{uncal}}$ are the calibrated and uncalibrated power spectra.

Multiplicative biases are dominantly caused by the effect of pixel noise in the galaxy images, resulting in a noise bias in maximum-likelihood shape estimators~\citep[e.g.][]{Bernstein2002, Melchior2012, Refregier2012, Miller2013}. They can also emerge from several other origins including disparities between the galaxy light profiles and the models used to describe them, causing a model bias~\citep[e.g.][]{Voigt2010, Melchior2010} or as a result of the quantities chosen for source selection, causing selection biases~\citep[e.g.][]{Jarvis2016}. The simulations used for the optical shear calibration here do not capture the effects of selection bias because the simulated sources are sampled from the weak lensing catalogue itself.

The two sets of component biases, $m_1$, $c_1$ and $m_2$, $c_2$ were measured separately by first holding $\gamma_2^{\rm sim} = 0$ and applying $\gamma_1^{\rm sim}$ in increments of 0.001 between 0 and 0.1, then repeating the process while keeping $\gamma_1^{\rm sim} = 0$ and incrementing $\gamma_2^{\rm sim}$. For each increment of $\gamma_{1,2}^{\rm sim}$, we randomly select 2,000 of the 243,852 sources from the optical weak lensing catalogue and draw their profiles with the widely used~\textsc{GalSim} image simulation package~\footnote{\url{https://github.com/GalSim-developers/GalSim/}}~\citep{Rowe2015}. The galaxies were drawn as two-component S\'{e}rsic profiles, each with a de Vaucouleurs bulge and an exponential disc (equation~\ref{eq:sersic_profile}) and distorted using the measured $\epsilon_{1,2}$ measurements from the weak lensing catalogue. For each source, the same measured radii and integrated flux combinations were incorporated into the simulated image to mimic the real data. Once drawn, the galaxies were individually given random orientations to remove any residual shear signal, and then all sheared with the same $\gamma_{1,2}^{\rm sim}$ values.

Each galaxy image was convolved with a randomly-selected PSF from the publicly available {\it HST}-ACS PSF sample from COSMOS observations~\footnote{\url{http://great3.jb.man.ac.uk/leaderboard/data/public/COSMOS_25.2_training_sample.tar.gz}}, and drawn with the pixel scale of the {\it HST}-ACS observations of 0.03 arcsec pixel$^{-1}$. Pixel noise was added to the images by randomly sampling a Gaussian distribution with a standard deviation of 0.003 ADU pixel$^{-1}$ measured from the original {\it HST}-ACS image tiles.

Finally, the galaxy images were cut to square stamps of side length 40 pixels to mimic the procedure in~\cite{Tunbridge2016} and the profiles were measured using~\textsc{im3shape} with the same PSF used to convolve the galaxy image. These measurements used the same~\textsc{im3shape} input parameters used to create the optical weak lensing catalogue described in~\cite{Tunbridge2016} including using a two-component de Vaucouleurs bulge plus exponential disc S\'{e}rsic model.

Each set of 2,000 measured $\epsilon_{1,2}$ parameters were used to estimate $\gamma_{1,2}^{\rm meas}$ values following equation~(\ref{eq:avg_shear}) with equal weightings for each galaxy. The calibration was propagated through the different input $\gamma_{1,2}^{\rm sim}$ values, and $m_i$ and $c_i$ were extracted by applying a least-squares linear fit with equation~(\ref{eq:shear_calibration}).

The multiplicative and additive biases we measure are summarized in Table~\ref{tab:calibration_measurements}. All the biases are consistent with zero, and used with equation~(\ref{eq:shear_calibration_Cl}), our values for $m_{1,2}$ would increase the power spectrum estimates by ($2.2 \pm 2.9$)\%. Since this correction is both negligible given the size of our error bars (left panel of Fig.~\ref{fig:GRF_powerSpectra_EE}) and consistent with no correction, we choose not to apply a calibration to our optical shear measurements.

\begin{table}
	\centering
	\caption{Multiplicative, $m_i$ and additive, $c_i$ biases for the optical weak lensing catalogue shear calibration procedure, with reference to equation~(\ref{eq:shear_calibration}). When extracting $m_1$ and $c_1$ with $\gamma_2^{\rm sim} = 0$, we measure $c_2 = (4 \pm 6) \times 10^{-4}$ and when extracting $m_2$ and $c_2$ with $\gamma_1^{\rm sim} = 0$, we measure $c_1 = (7 \pm 6) \times 10^{-4}$.}
	\label{tab:calibration_measurements}
	\begin{tabular}{cccc} 
		\hline
		$m_1$ & $c_1$ & $m_2$ & $c_2$ \\
		\hline
		$-0.020 \pm 0.021$ & $-0.001 \pm 0.001$ & $-0.002 \pm 0.019$ & $-0.001 \pm 0.001$ \\
		\hline
	\end{tabular}
\end{table}

\subsection{Shear Power Spectra Results}
\label{sec:radopcrosspowspec}
\begin{figure*}
	\includegraphics[width=\textwidth]{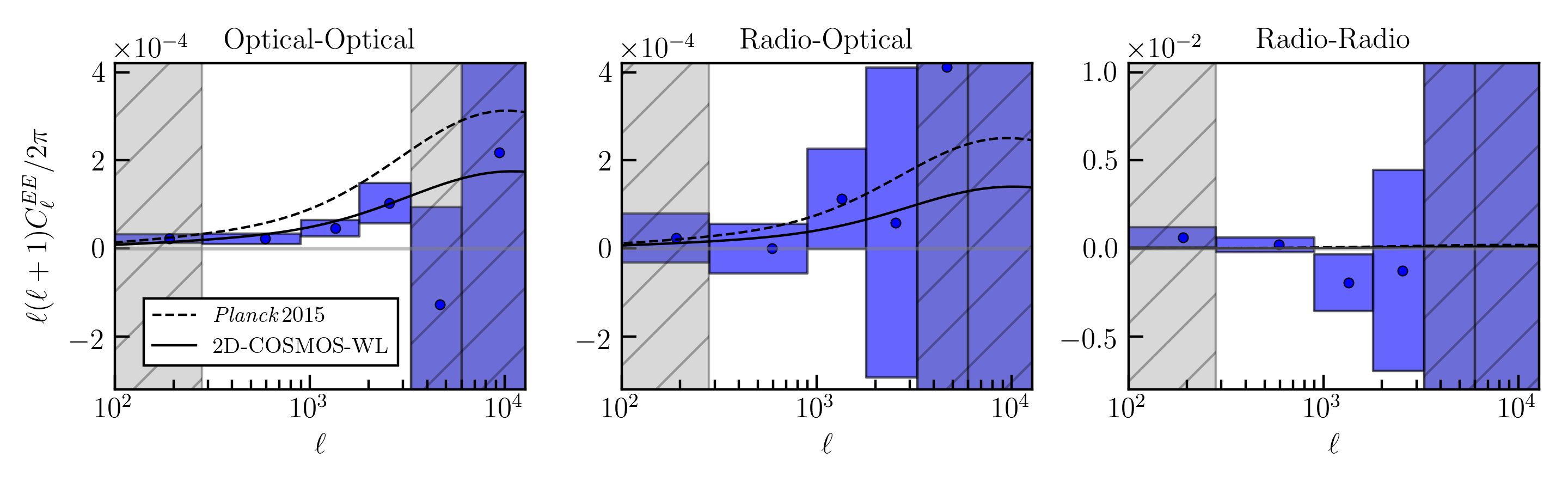}
    \caption{COSMOS $E$-mode power spectra. From left to right: optical-optical, radio-optical and radio-radio. Note the different vertical axis scale of the radio-radio power spectrum to accommodate the large error bars. Detection significances, signal-to-noise ratios and reduced chi-squared values were shown in Table~\ref{tab:powerspec_stats} and the parameters of the model spectra were shown in Table~\ref{tab:cosmologymodel}. The model spectra use the full redshift distribution given in equation~(\ref{eq:redshiftdist}) and have redshift combinations: $z_{\rm med}^{\rm O} \times z_{\rm med}^{\rm O}$, $z_{\rm med}^{\rm R} \times z_{\rm med}^{\rm O}$ and $z_{\rm med}^{\rm R} \times z_{\rm med}^{\rm R}$ from left to right, where $z_{\rm med}^{\rm O} = 1.26$ and $z_{\rm med}^{\rm R} = 1.00$. The error bars for these plots were generated using mock datasets, as described in Section~\ref{sec:GRFclones}.}
    \label{fig:EEpowerspectra_3panels_loglinear}
\end{figure*}

\begin{figure*}
	\includegraphics[width=\textwidth]{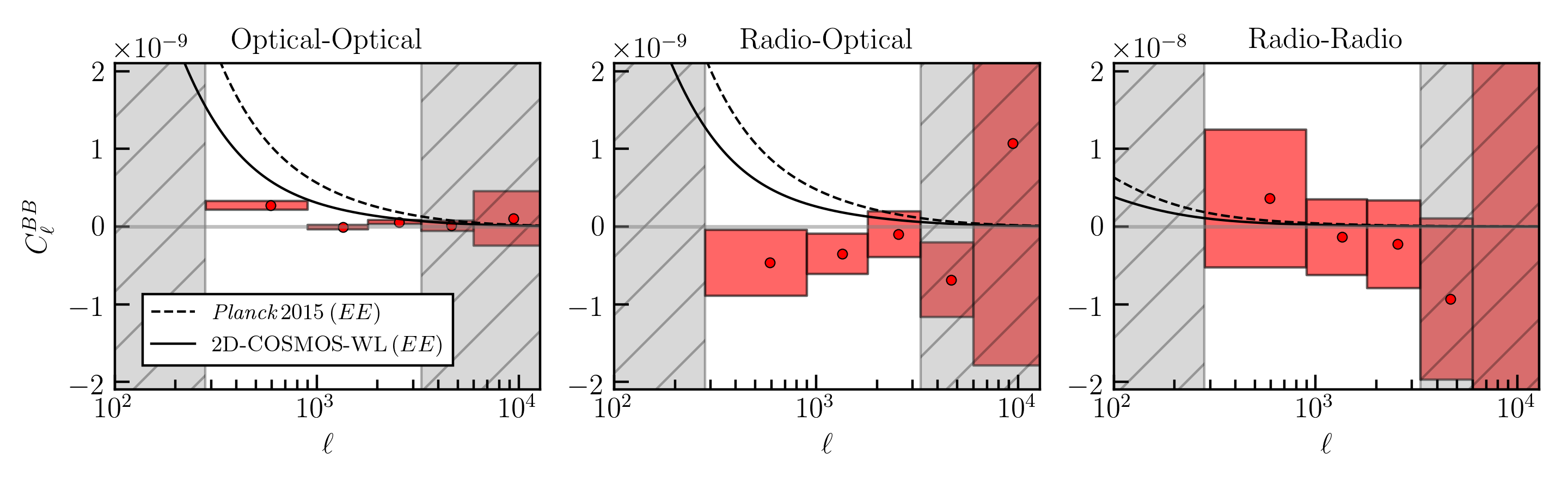}
    \caption{Same as Fig.~\ref{fig:EEpowerspectra_3panels_loglinear} but for COSMOS $B$-mode power spectra. Note the different band power and vertical axis scaling in comparison to Fig.~\ref{fig:EEpowerspectra_3panels_loglinear} and the different vertical axis scale of the radio-radio spectrum. The models shown are the same theoretical $E$-mode spectra shown in Fig.~\ref{fig:EEpowerspectra_3panels_loglinear}.}
    \label{fig:BBpowerspectra_3panels}
\end{figure*}

\begin{figure*}
	\includegraphics[width=\textwidth]{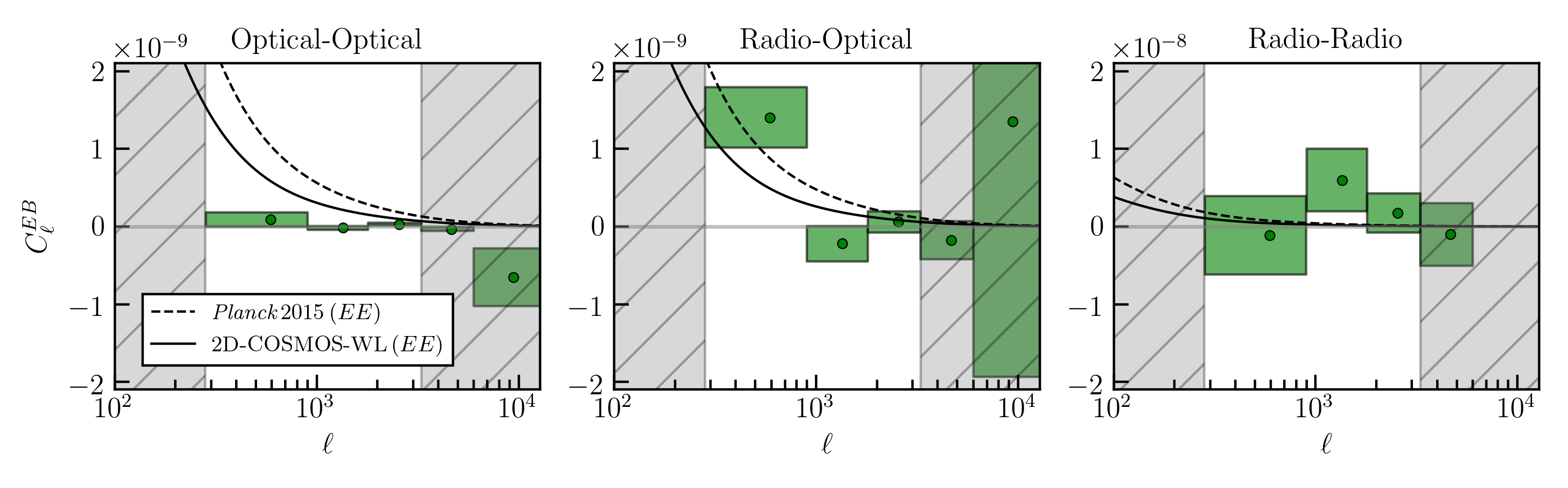}
    \caption{Same as Fig.~\ref{fig:BBpowerspectra_3panels} but for COSMOS $EB$ cross-correlation power spectra.}
    \label{fig:EBpowerspectra_3panels}
\end{figure*}

Having demonstrated the power spectrum recovery on the mock datasets and used these to estimate the uncertainties, we are now in a position to examine the power spectra recovered from the real data. These are shown in Figs.~\ref{fig:EEpowerspectra_3panels_loglinear},~\ref{fig:BBpowerspectra_3panels} and~\ref{fig:EBpowerspectra_3panels} for the $C_\ell^{EE}, C_\ell^{BB}$ and $C_\ell^{EB}$ power spectra respectively. 

The left-hand panel of Fig.~\ref{fig:EEpowerspectra_3panels_loglinear} shows the $E$-mode power spectrum that we measure from the {\it HST}-ACS optical data. Although these data have been used for previous weak lensing analyses~\citep{Massey2007, Schrabback2010}, our analysis represents the first direct measurement of the cosmic shear power spectrum from these data. Both of the previous analyses used real-space two-point statistics (i.e. the correlation functions and/or aperture mass dispersion) to quantify the lensing signal. While the real-space and Fourier-space two-point statistics should capture the same cosmological information, it is nevertheless both noteworthy and reassuring that the cosmic shear power spectrum that we recover from the {\it HST}-ACS data is in excellent agreement with the best-fitting $\Lambda$CDM cosmological model from the \cite{Schrabback2010} 2D correlation function analysis. Whilst the best-fitting model from the 2D analysis in \cite{Schrabback2010} favours a lower value for the $S_{8} = \sigma_{8}(\Omega_{\rm m}/0.3)^{0.5}$ parameter than the {\it Planck} $\Lambda$CDM model, this is very likely attributable to cosmic variance within this small field and hence should not be used to directly assess any levels of tension with larger surveys such as {\it Planck} or more recent weak lensing surveys~\citep[e.g.][]{DES2017, Hildebrandt2018, Hikage2019}.

The centre panel of Fig.~\ref{fig:EEpowerspectra_3panels_loglinear} shows our measurement of the radio-optical cross-correlation power spectrum, $C_\ell^{\rm RO}$. While the errors on this measurement are still relatively large due to the small number density of galaxies in the radio catalogue, our result is nevertheless also consistent with the 2D-COSMOS-WL cosmological model. However, this measurement is also consistent with the {\it Planck} best-fitting model, and indeed with zero signal. The right-hand panel of Fig.~\ref{fig:EEpowerspectra_3panels_loglinear} shows the constraints we obtain on the shear power spectrum from the radio data alone and illustrates the current limitations of weak lensing with radio surveys. (Note the expanded scale on the vertical axis for this plot compared to the other two panels in the figure.) The low number density of galaxies for which we are able to measure radio shapes means that the radio data on its own has essentially no constraining power. 

The results depicted in Figs.~\ref{fig:EEpowerspectra_3panels_loglinear},~\ref{fig:BBpowerspectra_3panels} and~\ref{fig:EBpowerspectra_3panels} are further quantified in Table~\ref{tab:powerspec_stats}. For each $C_\ell^{EE}$ spectrum, this table lists the significance with which a lensing signal is detected (the ``detection significance") and the signal-to-noise ratio of the measurement, accounting for both noise and cosmic variance. 
The detection significances were calculated as
\begin{equation}
    D = \sqrt[]{\sum_b \left ( \frac{\widehat{P}_b^{EE}}{\sigma_b^\prime} \right )^2},
	\label{eq:detectionsignificance}
\end{equation}
where the index $b$ runs over the number of included bands (bands 2, 3  and 4), $\widehat{P}_b^{EE}$ are the $E$-mode band powers measured from the data, and $\sigma_b^\prime$ are the uncertainties on the $E$-mode band powers {\it excluding cosmic variance}. We estimate $\sigma_b^\prime$ from the standard deviation of the $B$-mode band powers~\footnote{Since there is no input $B$-mode signal in the GRF simulations, the run-to-run scatter of the recovered $B$-mode band power estimates is free of cosmic variance.} recovered from the mock datasets,
\begin{equation}
    \sigma_b^\prime = \sqrt[]{\left < \left (P_b^{BB} \right )^2 \right > - \left < P_b^{BB} \right > ^2},
	\label{eq:detection_noise_error}
\end{equation}
where the angled brackets here denote an average over the mock datasets. Similarly, the signal-to-noise ratios were calculated according to
\begin{equation}
    S = \sqrt[]{\sum_b \left ( \frac{\widehat{P}_b^{EE}}{\sigma_b} \right )^2},
	\label{eq:snr_powerspec}
\end{equation}
where the uncertainties, $\sigma_b$ now include both measurement noise and cosmic variance. We estimate $\sigma_b$ from the standard deviation of the $E$-mode band powers recovered from the mock datasets:
\begin{equation}
    \sigma_b = \sqrt[]{\left < \left (P_b^{EE} \right )^2 \right > - \left < P_b^{EE} \right > ^2}.
	\label{eq:snr_noise_error}
\end{equation}

In addition, for all of the recovered spectra, Table~\ref{tab:powerspec_stats} also reports reduced chi-squared values, calculated with reference to both the 2D-COSMOS-WL best-fitting model, and with reference to a null signal. The reduced chi-squared values were calculated as
\begin{equation}
    \chi^2_{\rm red} = \frac{1}{\nu} \sum_{b} \left ( \frac{\widehat{P}_b - P_{b}^{\rm th}}{\sigma_{b}} \right ) ^2,
	\label{eq:chi_squared_cov}
\end{equation}
where $\sigma_{b}$ represents the same quantity as in equation~(\ref{eq:snr_noise_error}) and the index $b$ runs over the number of included bands, $\nu$. When comparing to the null signal, the model values, $P_{b}^{\rm th}$, were set to zero. When comparing to the best-fitting 2D-COSMOS-WL model, the model values were set to the band power expectation values for that model, calculated according to equation~(\ref{eq:band_expectation}).

The results reported in Table~\ref{tab:powerspec_stats} further demonstrate that the optical lensing signal ($C_\ell^{\rm OO}$) is detected with very high significance while the other channels ($C_\ell^{\rm RO}$ \& $C_\ell^{\rm RR}$) do not deliver a significant detection. The results also demonstrate that the $B$-mode and $EB$ cross-correlation band powers are much more consistent with a null signal than they are with the 2D-COSMOS-WL best-fitting shear signal.

\begin{table}
	\centering
	\caption{Detection significances, $D$, signal-to-noise ratios, $S$, and reduced chi-squared values, $\chi_\mathrm{red}^2$, for the band power measurements shown in Figs.~\ref{fig:EEpowerspectra_3panels_loglinear},~\ref{fig:BBpowerspectra_3panels} and~\ref{fig:EBpowerspectra_3panels}. The reduced chi-squared values were calculated using both the band power-convolved 2D-COSMOS-WL model and the zero line. The detection significances and signal-to-noise ratios shown in parentheses are those calculated using the mean signal recovered from the mock datasets shown in Figs.~\ref{fig:GRF_powerSpectra_EE},~\ref{fig:GRF_powerSpectra_BB} and~\ref{fig:GRF_powerSpectra_EB}. All values were determined by only using the measurements of bands 2, 3 and 4 (Table~\ref{tab:bandpowerselection}).}
	\label{tab:powerspec_stats}
	\begin{tabular}{c c c c c} 
		\hline
        Spectrum & $D$ & $S$ & \multicolumn{2}{c}{$\chi_{\mathrm{red}}^2$} \\
        & & & To & To \\
        & & & $E$-mode & Zero \\
        & & & Model & Line \\
        \hline
        $\mathrm{OpOp}^{EE}$ & 9.80 (12.03) & 3.88 (4.42) & 0.39 & \color{mygrey}{5.01} \\
        $\mathrm{RadOp}^{EE}$ & 1.50 (1.65) & 1.00 (0.82) & 0.17 & \color{mygrey}{0.34} \\
        $\mathrm{RadRad}^{EE}$ & 1.45 (0.26) & 1.32 (0.28) & 0.59 & \color{mygrey}{0.58} \\
        $\mathrm{OpOp}^{BB}$ & - & - & \color{mygrey}{26.30} & 10.00 \\
        $\mathrm{RadOp}^{BB}$ & - & - & \color{mygrey}{2.99} & 1.04 \\
        $\mathrm{RadRad}^{BB}$ & - & - & \color{mygrey}{0.13} & 0.14 \\
        $\mathrm{OpOp}^{EB}$ & - & - & \color{mygrey}{35.44} & 1.34 \\
        $\mathrm{RadOp}^{EB}$ & - & - & \color{mygrey}{2.98} & 4.76 \\
        $\mathrm{RadRad}^{EB}$ & - & - & \color{mygrey}{0.88} & 0.91 \\
		\hline
	\end{tabular}
\end{table}

\subsection{Cross-Power Spectrum Noise Considerations}
\label{sec:crossnoiseconsiderations}

As discussed in Section~\ref{sec:crosscorrelationnoise}, the correlated shape noise term associated with a cross-power spectrum is dependent on the amount of sources common to both maps being analysed and the shape correlation between those sources.

For tomographic weak lensing analyses in a single waveband, such as in \cite{Kohlinger2017}, this term is zero, since there is no overlap of sources after binning by redshift. This is not the case here, where we have overlapping redshift bins. As was shown in Section~\ref{sec:shapecomparison}, specifically Fig.~\ref{fig:hist_separations}, there are 1,078 sources common to both the 3 GHz VLA and {\it HST}-ACS weak lensing catalogues, giving a matching percentage of 53\% of the radio and 0.44\% of the optical sources. Hence, the cross-power noise term is non-zero.

All the figures and values in Section~\ref{sec:powerspectra} extracted for the $E$-mode optical-optical and radio-radio auto-power spectra have correlated shape noise accounted for according to equation~(\ref{eq:noiseauto}). However, the radio-optical cross-power spectra do not have any correlated shape noise subtracted from them. This noise should be calculated according to equation~(\ref{eq:noisecrossfull}). Note that this only applies to the real COSMOS data radio-optical spectrum in Fig.~\ref{fig:EEpowerspectra_3panels_loglinear}, and not Fig.~\ref{fig:GRF_powerSpectra_EE}, since the 100 mock datasets had randomly-assigned galaxy shapes from the two catalogues.

\begin{figure}
	\includegraphics[width=\columnwidth]{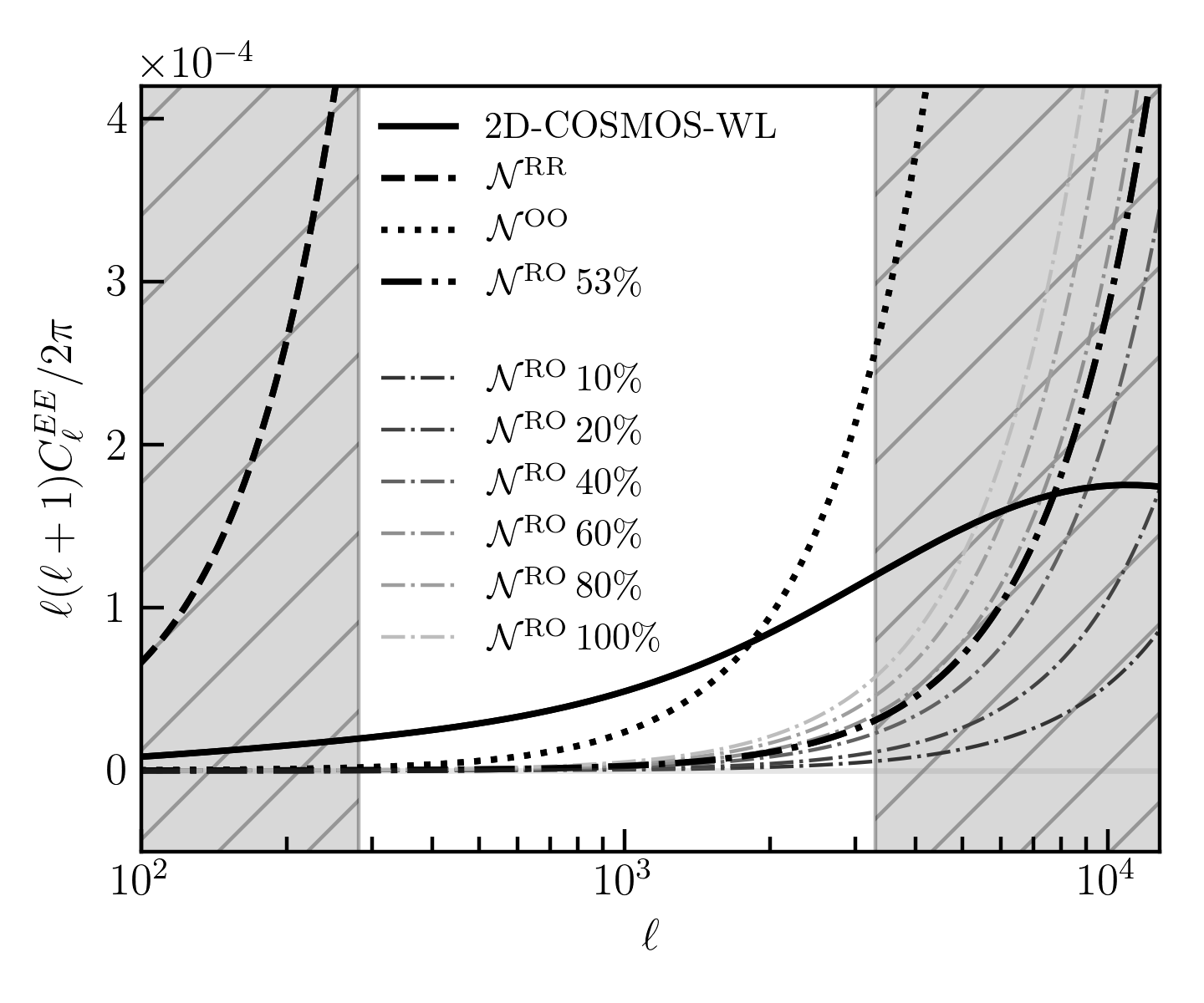}
    \caption{Noise power spectra calculated using equations~(\ref{eq:noiseauto}) and (\ref{eq:noisecrossfull}). The thick black curves represent estimates for the radio-radio, optical-optical and radio-optical correlated shape noise in the $E$-mode power spectra presented in Fig.~\ref{fig:EEpowerspectra_3panels_loglinear}. The thin grey curves represent estimates for the radio-optical noise using matchings between 10\% and 100\% of the radio catalogue with the same measured covariance of ${\rm cov}(\epsilon_{\rm R}, \epsilon_{\rm O}) = 0.016$. The hatched exclusion regions match with those in e.g. Fig.~\ref{fig:EEpowerspectra_3panels_loglinear}.}
    \label{fig:noise_spectra_linear}
\end{figure}

Fig.~\ref{fig:noise_spectra_linear} shows some estimates of the noise power spectra expected for each of the $E$-mode power spectra using equation~(\ref{eq:noiseauto}) for $\mathcal{N}^{\rm OO}$ and $\mathcal{N}^{\rm RR}$, and equation~(\ref{eq:noisecrossfull}) for $\mathcal{N}^{\rm RO}$ along with the source number densities in Table~\ref{tab:WLcatalogues}. For the $\mathcal{N}^{\rm RO}$ noise curve, the covariance term was calculated according to
\begin{equation}
    {\rm cov}(\epsilon_{\rm R}, \epsilon_{\rm O}) = \left < \left ( \epsilon_{\rm R} - \left < \epsilon_{\rm R} \right > \right ) \left ( \epsilon_{\rm O} - \left < \epsilon_{\rm O} \right > \right ) \right >,
	\label{eq:cov_er_eo}
\end{equation}
where the angled brackets denote the average over all 1,078 matched sources and both $\epsilon_1$ and $\epsilon_2$ components are included. For the 1,078 sources common to both the radio and optical catalogues, we measure a covariance of ${\rm cov}(\epsilon_{\rm R}, \epsilon_{\rm O}) = 0.016$.

As shown in Fig.~\ref{fig:noise_spectra_linear}, the $\mathcal{N}^{\rm OO}$ and $\mathcal{N}^{\rm RR}$ curves are significant in comparison to the 2D-COSMOS-WL theory curve. However, the $\mathcal{N}^{\rm RO}$ curve with the matching fraction of 53\% is negligible on the scales we probe in this work, justifying the decision to ignore the correlated shape noise in the estimation of the radio-optical $E$-mode power spectrum. To illustrate the possible spread on the cross-power spectrum noise curve, Fig.~\ref{fig:noise_spectra_linear} also shows some estimates around the $\mathcal{N}^{\rm RO} 53\%$ curve with matching fractions ranging from 10\% to 100\% of sources in the radio catalogue, using the same measured covariance. Again, we note that these curves are negligible on the scales we probe in this work, especially considering the size of the statistical error bars produced by the low source number density of the radio catalogue.

\section{Conclusions}
\label{sec:conclusions}
The weak lensing analysis of the 3 GHz VLA radio and {\it HST}-ACS optical surveys of the COSMOS field presented here demonstrated two correlations between the two datasets.

Firstly, the correlation between the intrinsic shapes of radio and optical galaxies was measured by comparing galaxy position angles. A Pearson correlation coefficient of $R_\alpha = 0.14 \pm 0.03$ was measured using 1,078 cross-matched sources, as illustrated in Fig.~\ref{fig:alpha_Comparison_Op_3GHz_2dhist}. This coefficient is a marked improvement on previous work which found much weaker, or non-existent correlations. The correlation gives insight into the emission processes associated with radio and optical galaxies and can be used to inform estimates of the correlated shape noise between two shear maps in the cross-power spectrum analysis, as discussed in Section~\ref{sec:crossnoiseconsiderations}.

Secondly, a cross-correlation power spectrum between the radio and optical shear fields was extracted, along with optical-optical and radio-radio auto-power spectra. The optical-optical $E$-mode auto-power spectrum gave a detection significance of 9.80$\sigma$ and was in excellent agreement with the best-fitting $\Lambda$CDM cosmological model from the \cite{Schrabback2010} 2D correlation function analysis. Our radio-optical $E$-mode cross-power spectrum did not show a significant detection (1.50$\sigma$), although the measurement was determined to be more consistent with the theoretical expectation cross-power spectrum than with a null signal. The results shown in Table~\ref{tab:powerspec_stats} and the centre panels of Figs.~\ref{fig:EEpowerspectra_3panels_loglinear},~\ref{fig:BBpowerspectra_3panels} and~\ref{fig:EBpowerspectra_3panels} add weight to the argument that increasing source number densities in radio experiments will allow for significant detections of radio-optical and radio-radio shear power spectra. Also, Fig.~\ref{fig:noise_spectra_linear} shows that the correlated shape noise power spectra associated with the cross-correlation weak lensing analysis presented here is significantly diminished for reasonable expectations of overlapping source populations.

In measuring the shapes of the sources in the radio catalogue, we highlighted the issues associated with image-based shape measurement for radio observations including the nonlinear deconvolution step in imaging, and the presence of correlated noise between sources. In this work these issues were sub-dominant to the statistical uncertainties caused by the small number of sources in the radio catalogue. However, such issues will need to be addressed for future radio weak lensing surveys.

Pathfinder surveys such as this 3 GHz VLA COSMOS data and SuperCLASS will help lead the field on to radio weak lensing with the SKA, which will be the first cosmologically competitive radio weak lensing survey. Ultimately, this will help to achieve cosmological parameter constraints that are more robust against instrumental and astrophysical systematic biases.

\section*{Acknowledgements}

We thank Vernesa Smol\v{c}i\'{c} and Mladen Novak for providing the 3 GHz VLA COSMOS real-space image used for this paper and Ben Tunbridge for providing the {\it HST}-ACS optical shape catalogue. The 3 GHz VLA COSMOS data is based on observations with the National Radio Astronomy Observatory, a facility of the National Science Foundation operated under cooperative agreement by Associated Universities, Inc. We further thank Fabian K\"{o}hlinger for useful correspondence regarding the use of the power spectrum estimator.

TH is supported by a Science and Technology Facilities Council studentship. IH acknowledges the support of the Beecroft Trust and from the European Research Council in the form of a Consolidator Grant with number 681431.




\bibliographystyle{mnras}
\bibliography{COSMOS_3GHz_refs}







\bsp	
\label{lastpage}
\end{document}